%% file: paper2.tex
\documentclass[a4paper, 11pt]{article}
\usepackage{amssymb}
\usepackage{epsfig}
\usepackage{enumerate}
\psfigdriver{dvips}

\begin{document}
\include{journals}

\newcommand{\uJy}[0]{\ensuremath{\mu\textrm{Jy}}}
\newcommand{\Lya}[0]{\ensuremath{\textrm{Ly--}\alpha}}

\title{Deep optical imaging of the field of PC1643+4631A\&B,\\ II:
Estimating the colours and redshifts of faint galaxies}

\author{Garret Cotter$^{\! 1}$, Toby Haynes$^{\! 1}$, \\ Joanne
C. Baker$^{\! 2}$, Michael E. Jones$^{\! 1}$, Richard Saunders$^{\! 1}$\\ \\ 
$^1$ Astrophysics, Cavendish Laboratory,\\ 
Madingley Road, Cambridge, CB3 0HE, UK\\
$^2$ Astronomy Department, University of California, \\
Berkeley CA 94720, USA
}

\maketitle

\begin{abstract}

In an investigation of the cause of the cosmic microwave background
decrement in the field of the $z = 3.8$ quasar pair PC1643+4631, we
have carried out a study to photometrically estimate the redshifts of
galaxies in deep multi-colour optical images of the field taken with
the WHT.  

To examine the possibility that a massive cluster of galaxies lies in
the field, we have attempted to recover simulated galaxies with
intrinsic colours matching those of the model galaxies used in the
photometric redshift estimation. We find that when such model galaxies
are added to our images, there is considerable scatter of the
recovered galaxy redshifts away from the model value; this scatter is
larger than that expected from photometric errors and is the result of
confusion, simply due to ground-based seeing, between objects in the
field.

We have also compared the likely efficiency of the photometric
redshift technique against the colour criteria used to select $z
\gtrsim 3$ galaxies via the strong colour signature of the Lyman-limit
break. We find that these techniques may significantly underestimate
the true surface density of $z \gtrsim 3$, due to confusion between
the high--redshift galaxies and other objects near the line of
sight. We argue that the actual surface density of $z\approx3$
galaxies may be as much as 6 times greater than that estimated by
previous ground-based studies, and note that this conclusion is
consistent with the surface density of high--redshift objects found in
the HDF. Finally, we conclude that all ground--based deep field
surveys are inevitably affected by confusion, and note that reducing
the effective seeing in ground--based images will be of paramount
importance in observing the distant universe.

\end{abstract}

\section{Introduction}

\label{sec:print}

The deep five--colour photometry (Paper I) of PC1643+4631 allow an
attempt to estimate the redshifts of objects in the field.  To make
the best use of our images, we must have an equivalent set of colours
of galaxies derived from either observations or models. Composite
galaxy spectra, such as those compiled by \cite{CWW} (hereafter known
as CWW), can be convolved with the telescope transmission functions
through each filter so that the colours of such a galaxy at various
redshifts can be modelled, providing us with a ``no--evolution mode''
set of colours. The original CWW spectra cover wavelengths from
1500\AA\ to 10000\AA , while the filters effectively sample from
3000\AA\ to 9000\AA , so such an approach would provide model colours
for $z<1$. Extending these spectra further into the UV allows models
for higher redshifts to be calculated and this can be achieved by
careful matching of UV spectra, eg from the \cite{KBC93} atlas, with
the blue end of the optical spectra. The different morphological types
of galaxies must also be taken into account -- spiral galaxies tend,
for example, to be considerably bluer than ellipticals -- so different
sample spectra must be compiled for each morphological group.

The main restriction encountered with these composite spectra is the
difficulty in adapting these spectra to account for evolution. This is
not really a problem at low redshifts, but to extend these models back
to the possible formation redshifts at $z_f=5$ one must include the
evolution of these galaxies. The best one can do at present is to use
the existing models, in which typically there is a burst of star
formation at some formation redshift, $z_f$, followed by exponentially
decreasing star-formation, and calculate the consequent colour
evolution.

\section{Modelling the galaxy colours over {$0<z<4$}}

The composite spectra from actual galaxies provide the most
representative colours for the low redshift galaxies in the field
(see, for example \cite{SG98}). At higher redshifts there are
significant differences between the colours derived from empirical and
simulated (in this case using the Bruzual \& Charlot (B\&C hereafter,
\cite{BC93}) code) spectra: this is the result of including evolution
in the simulated spectra. In order to achieve a consistent set of
colours for each morphological type for redshifts between 0 and 4, one
cannot simply rely on either the B\&C simulations or the CWW spectra
-- the former do not match the low--redshift galaxy spectra so well,
while the latter do not include any evolution and can not, in any
case, be used beyond $z\approx1.5$ since the UV end of the spectrum
starts to move out of the $U$ filter at this redshift.

\begin{figure} 
\begin{minipage}[hp]{0.5\linewidth}
\epsfig{figure=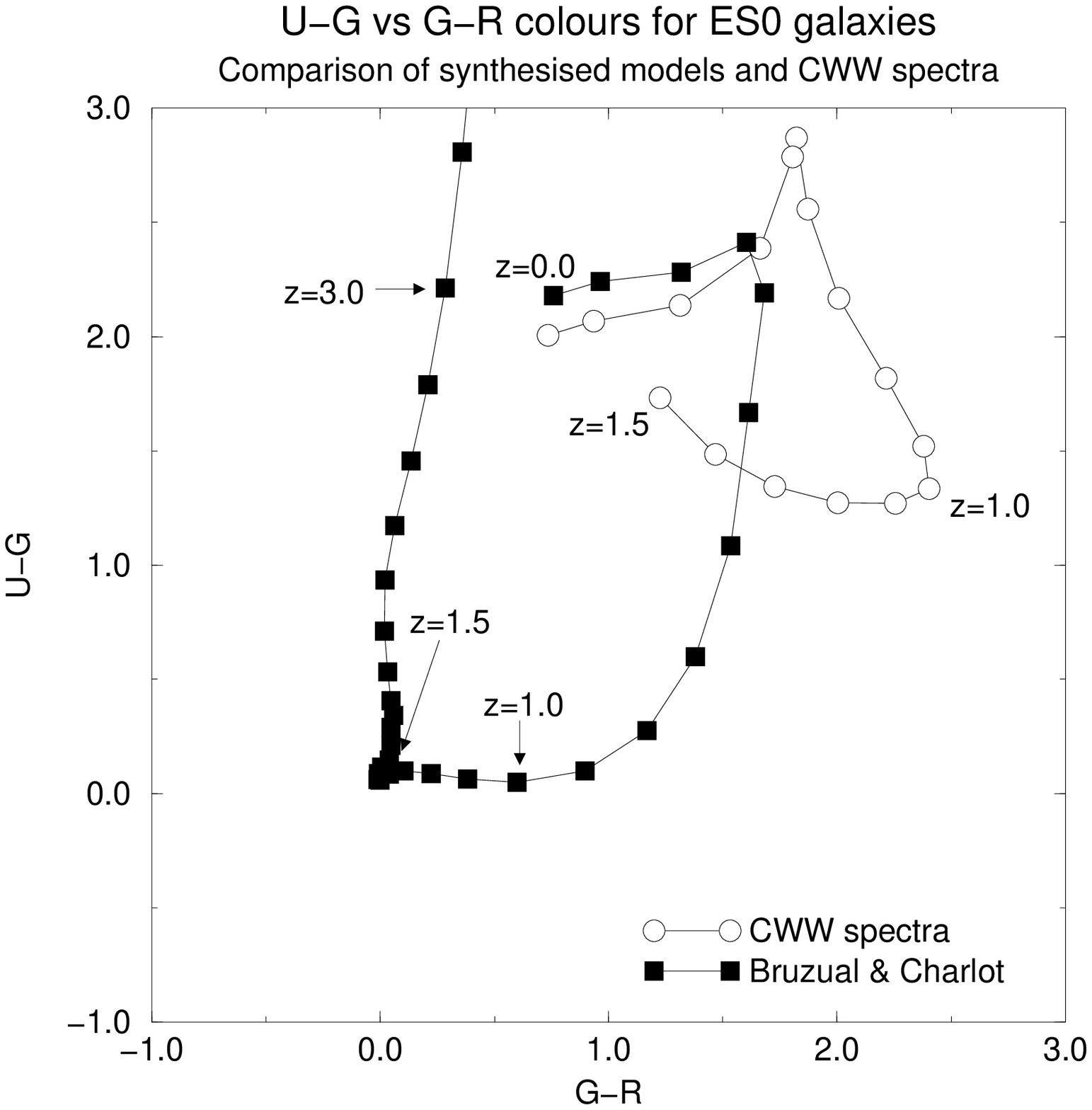,width=0.8\linewidth,angle=270}
\end{minipage}\hfill
\begin{minipage}[hp]{0.5\linewidth}
\epsfig{figure=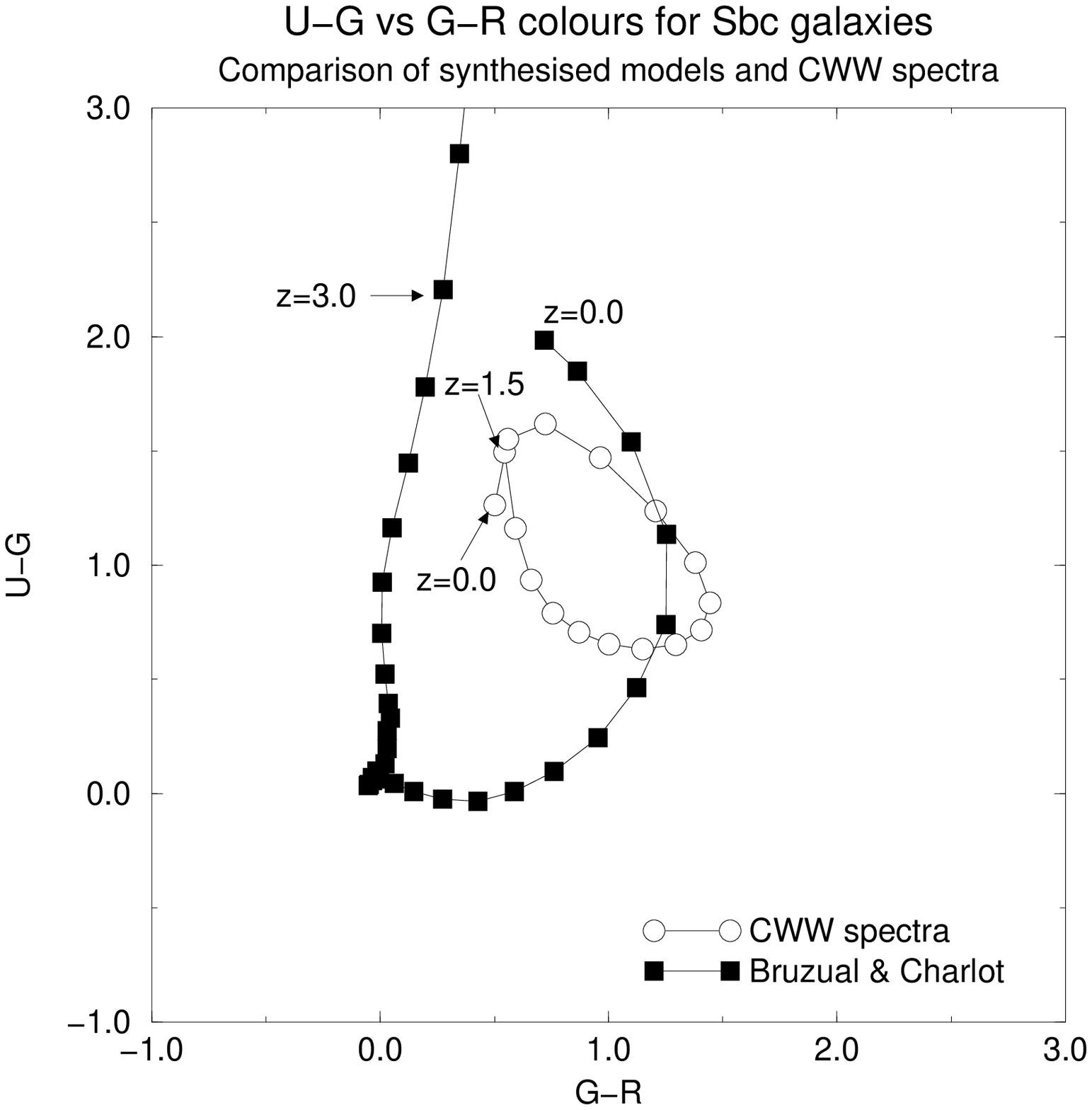,width=0.8\linewidth,angle=270}
\end{minipage}
\begin{minipage}[hp]{0.5\linewidth}
\epsfig{figure=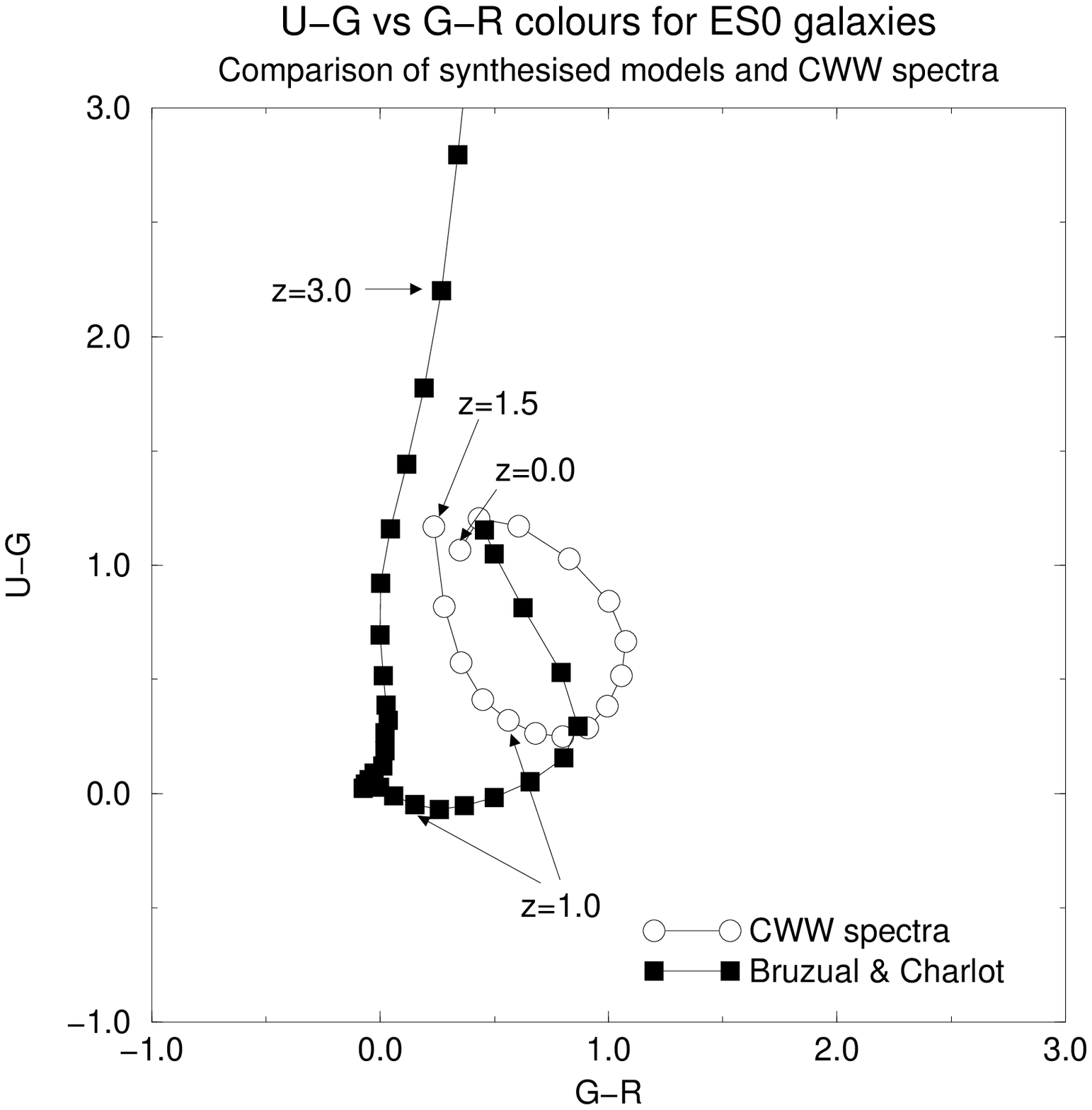,width=0.8\linewidth,angle=270}
\end{minipage}\hfill
\begin{minipage}[hp]{0.5\linewidth}
\epsfig{figure=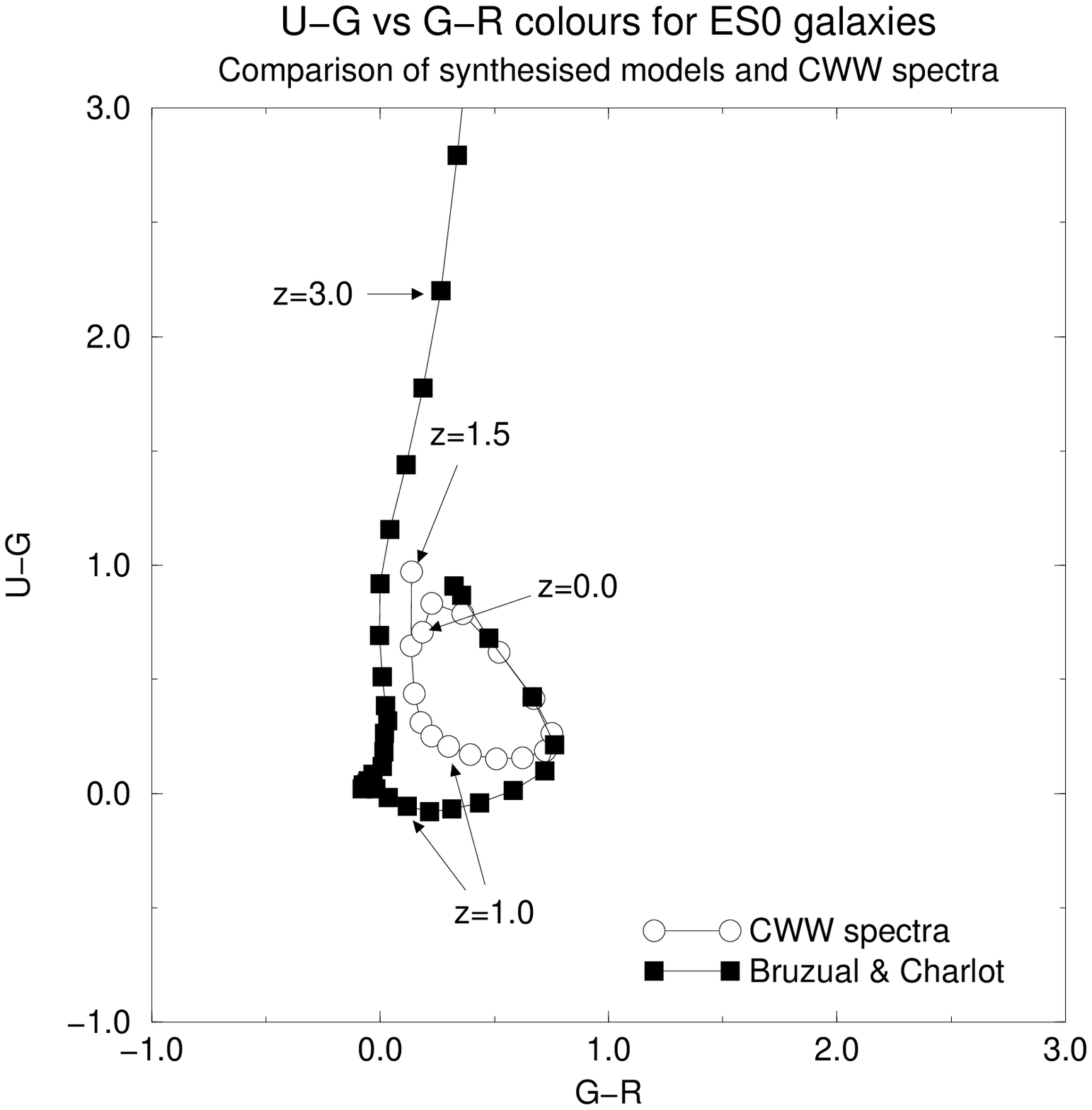,width=0.8\linewidth,angle=270}
\end{minipage}
\caption[Comparison of simulated and empirical spectra]
{Comparison of U-G and G-R colours based on simulated galaxy spectra
for four morphological types against the empirical galaxy spectra
taken from Coleman, Wu and Weedman\label{fig:galcomp}. The points are
at intervals of 0.1 in redshift}
\end{figure}

Colours derived from both sets of galaxy spectra are shown in
Figure~\ref{fig:galcomp}. Note that all the CWW spectra are much
redder in both $U-G$ and $G-R$ at $z=1.5$ than the B\&C spectra. At
this redshift, the star formation rate in the B\&C spectra is
producing considerably more flux in the UV than is seen in local
galaxies, consistent with the apparent star formation history (see,
eg. \cite{MPD98}). The result is that across the optical observing
band (3000\AA\ -- 8500\AA ), the spectra of star--forming galaxies are
close to flat spectrum for $1.2<z<2.3$. The success of the
Lyman--limit imaging searches for $z\gtrsim3$ galaxies using
B\&C--derived spectra indicates that at least some galaxies match
these simulated spectra at high redshift.

The approach adopted by Gwyn (\cite{SG98}) has changed since his
1996 paper (\cite{GH96}) on photometric redshifts in the Hubble Deep
Field (HDF). Previously, he relied on using spectra derived entirely
from B\&C simulations. Now, where HDF galaxies are found to be at
$z\le1.5$ using the B\&C spectra, he derives new photometric redshifts
from colours based on the CWW spectra and reports improved accuracy
when compared against spectroscopic redshifts. However, this results
in a large hiatus between the colours of a $z=1.6$, B\&C--derived
galaxy and a $z=1.5$, CWW--derived galaxy, which is unsatisfactory, as
can be seen in Figure~\ref{fig:galcomp}.

To attempt to make use of the CWW spectra to more accurately estimate
the colours of low--redshift galaxies, we have used the following
approach:\\ 

\begin{tabular}{rl}
   for $z\le0.5$, & galaxy colours are derived only from the \\
   	& CWW empirical spectra; \\

   for $0.5<z\le1.5$, 
	& galaxy colours are interpolated smoothly between the \\
   	& two sets of models, with colours at $z\sim0.6$ being \\
   	& closest to the CWW galaxy colours, and the colours at \\
   	& $z\sim1.5$ being closest to the B\&C colours; \\

   for $z>1.5$, & galaxy colours are derived only from the \\
   	& B\&C simulated spectra. \\

\end{tabular}
\smallskip

\begin{figure}
\begin{center}
\epsfig{figure=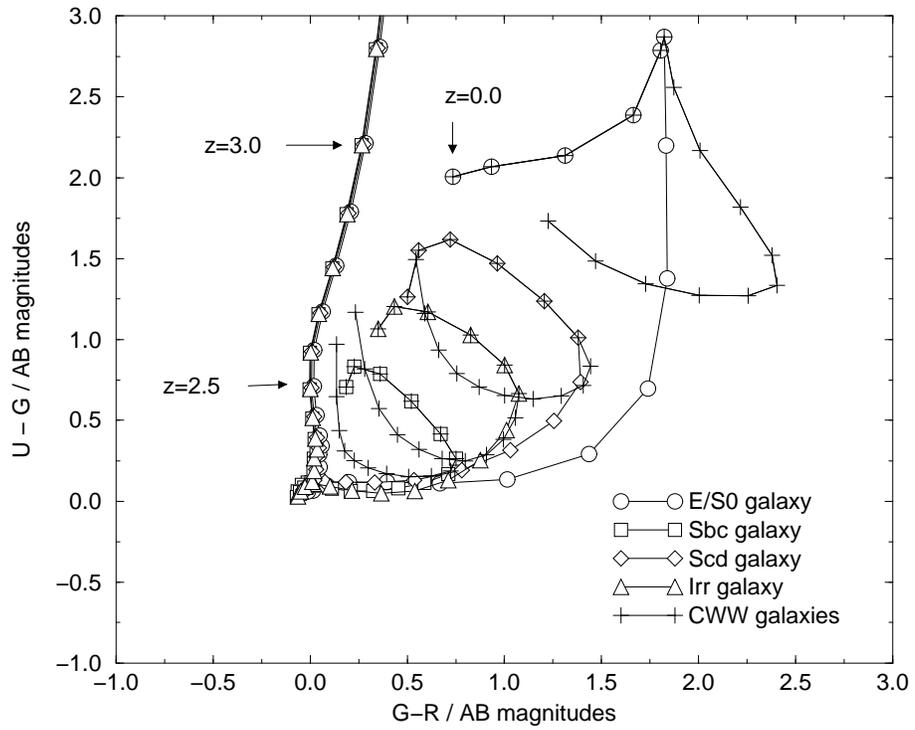,angle=270,width=0.95\linewidth}
\caption{Final model colours. Points shown are at intervals of 0.1 in redshift.
\label{fig:finalcols}
}
\end{center}
\end{figure}

The final galaxy models used are illustrated in
Figure~\ref{fig:finalcols}. The CWW spectra are over-plotted to show
the differences between the models.

Four morphological types were used, corresponding to E/S0, Sbc, Scd
and Irr, and spectra were produced at redshifts from 0.0 to 4.0 at 0.1
intervals. To cover the colour space more completely, these four models
were interpolated linearly between the model types and in redshift, to
create 13 models at redshift intervals of 0.025.

\section[Matching the model spectra]
{Matching the model spectra against the observations}

Since the number of counts received even in a small faint aperture is
large due to the plates being background limited, one can treat all
errors as gaussian about the mean flux. Fluxes (unnormalised) for the
model galaxies are derived from the model spectra by convolution with
the telescope response for each filter, including atmospheric
absorption and the effect of scattering within the telescope light
path.

To compare the models against the observed fluxes, we used a $\chi^2$
test, with

\begin{equation}
 \chi^{2} = \sum_i \Big(\frac {f\!_i - \alpha m_i}{\sigma_i}\Big)^2,
\end{equation}

where $f_i$ is the observed flux in filter $i$, $\sigma_i$ is the
error in observed flux in filter $i$, and $m_i$ is the model flux in
filter $i$. To obtain the optimum fit between the model and the observed fluxes,
we minimised $\chi^2$ with respect to $\alpha$.

However, given the range of models available, it is important to
determine the range of best--fit models which are consistent with the
observations, and for this we used a relative likelihood method. The
likelihood of obtaining an observed set of fluxes $f_i$ with errors
$\sigma_i$ from model fluxes $m_i$ (where $m_i$ is a function of
redshift, $z$, and morphological type, T) is

\begin{equation}
L(z,T) = \prod_i \exp \bigg[ -\frac{1}{2}\Big(\frac {f\!_i - \alpha
m_i(z,T)}{\sigma_i}\Big)^2 \bigg].
\end{equation}

This assumes that the errors in the observed fluxes are described
by a Gaussian distribution -- a good assumption. From this likelihood
distribution, the modal redshift can be determined by identifying the
model with the highest likelihood.

Since one can apportion a probability to each model, one can use this
to determine not only the most likely model, but also the mean
redshift and the range of acceptable redshifts which are consistent
with the observations within some relative confidence level. This is
important, since for any set of models the distribution of likely
models may not be mono--modal. We can use the difference between the
mean and modal redshifts as a simple discriminator to identify cases
where the redshift is not well constrained.

\section{Redshift distribution}

\subsection{Overview}
\label{sec:over}

\begin{figure}
\begin{center}
\epsfig{figure=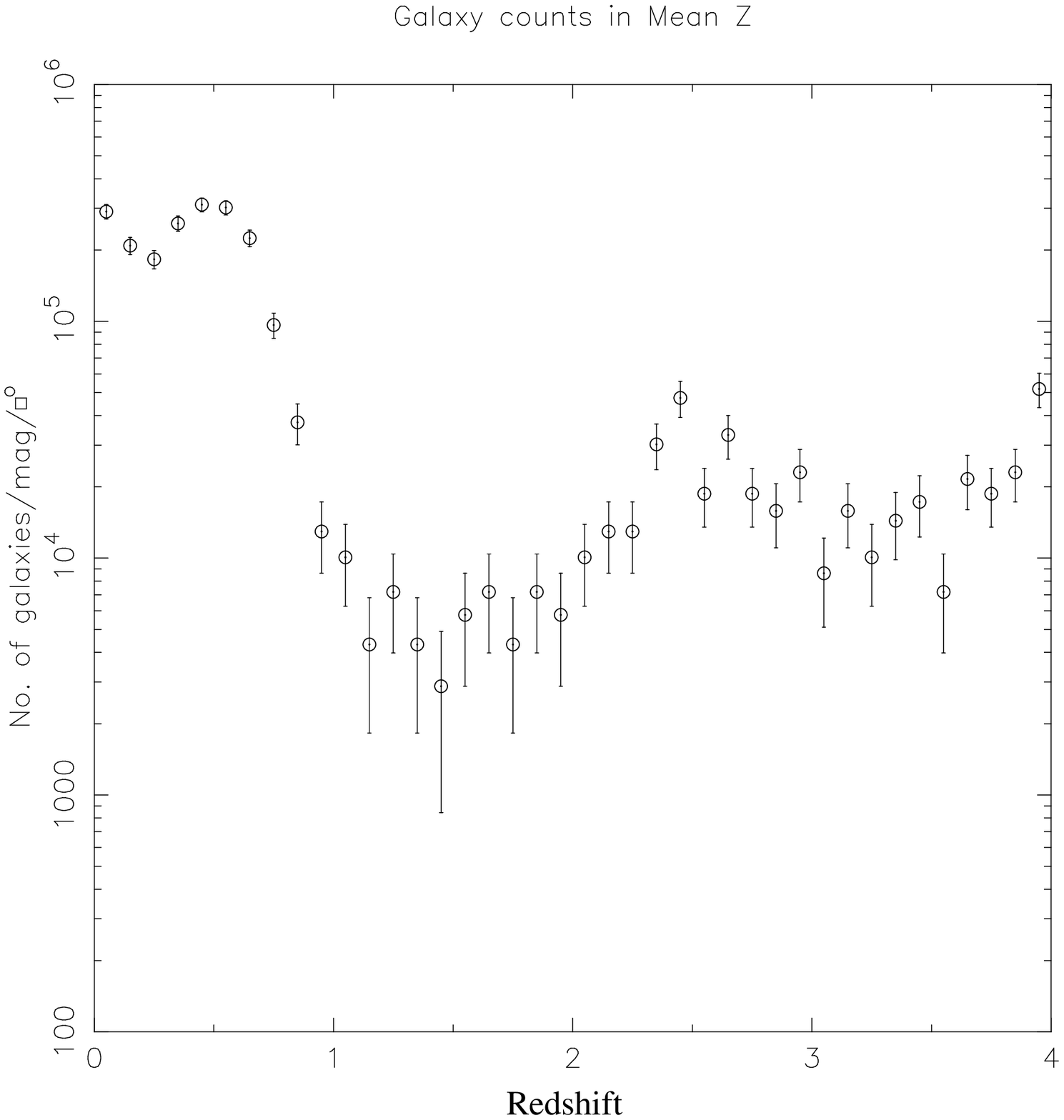,angle=0,width=0.6\linewidth}
\epsfig{figure=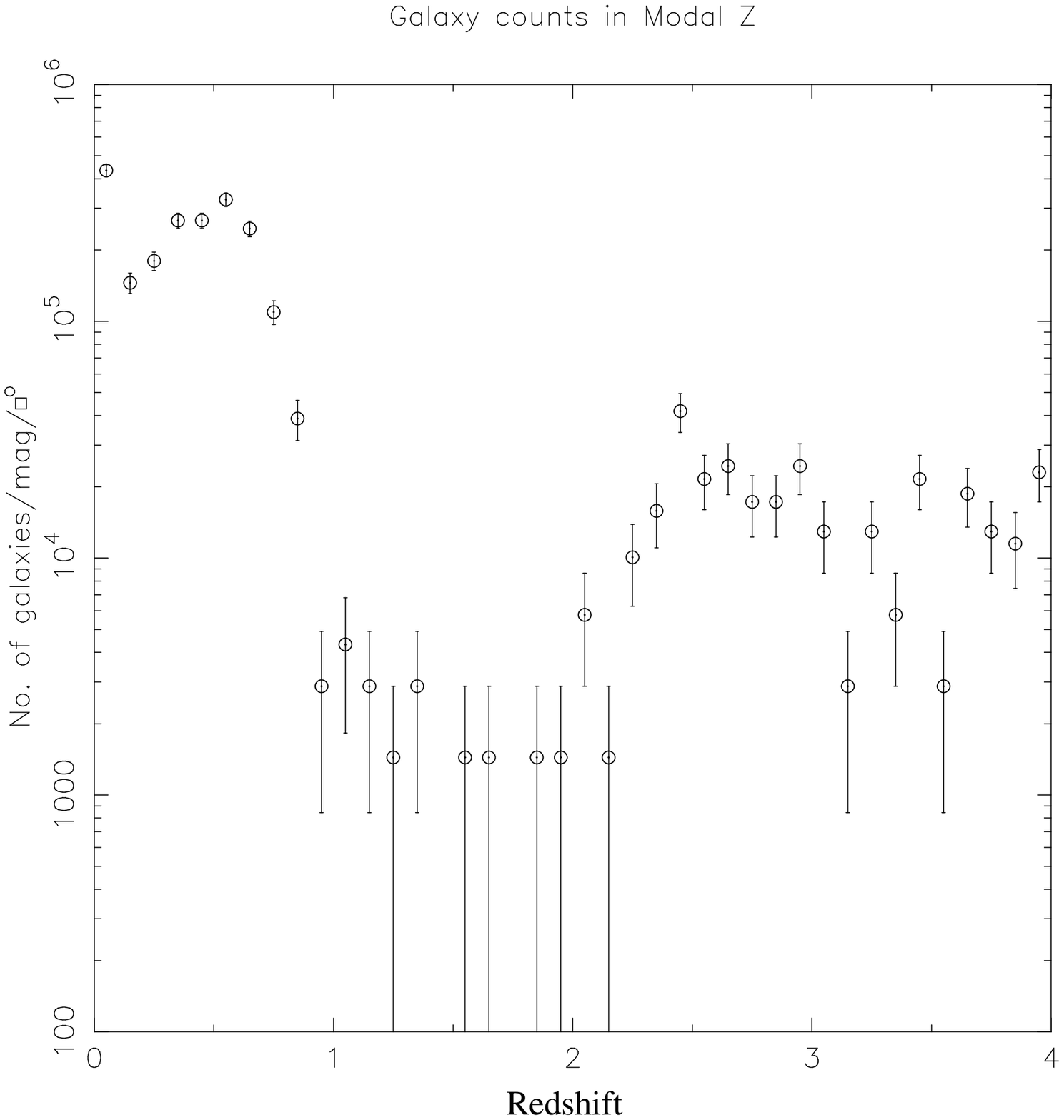,angle=0,width=0.6\linewidth}
\caption{Distributions based on photometric redshifts for all objects
with $R<26.0$
\label{fig:zhist}
}
\end{center}
\end{figure}

\begin{figure}
\begin{center}
\epsfig{figure=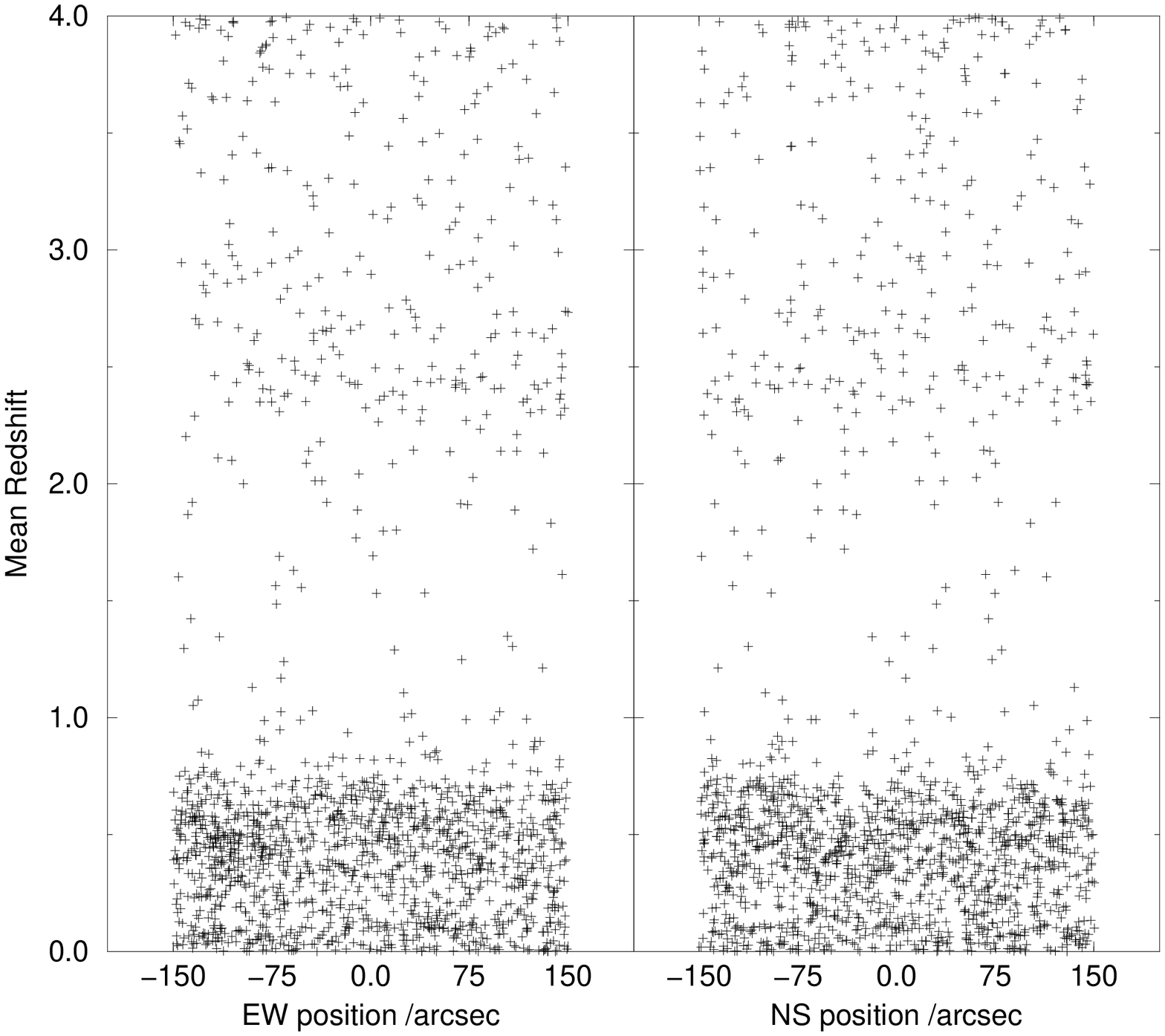,angle=270,width=0.7\linewidth}
\epsfig{figure=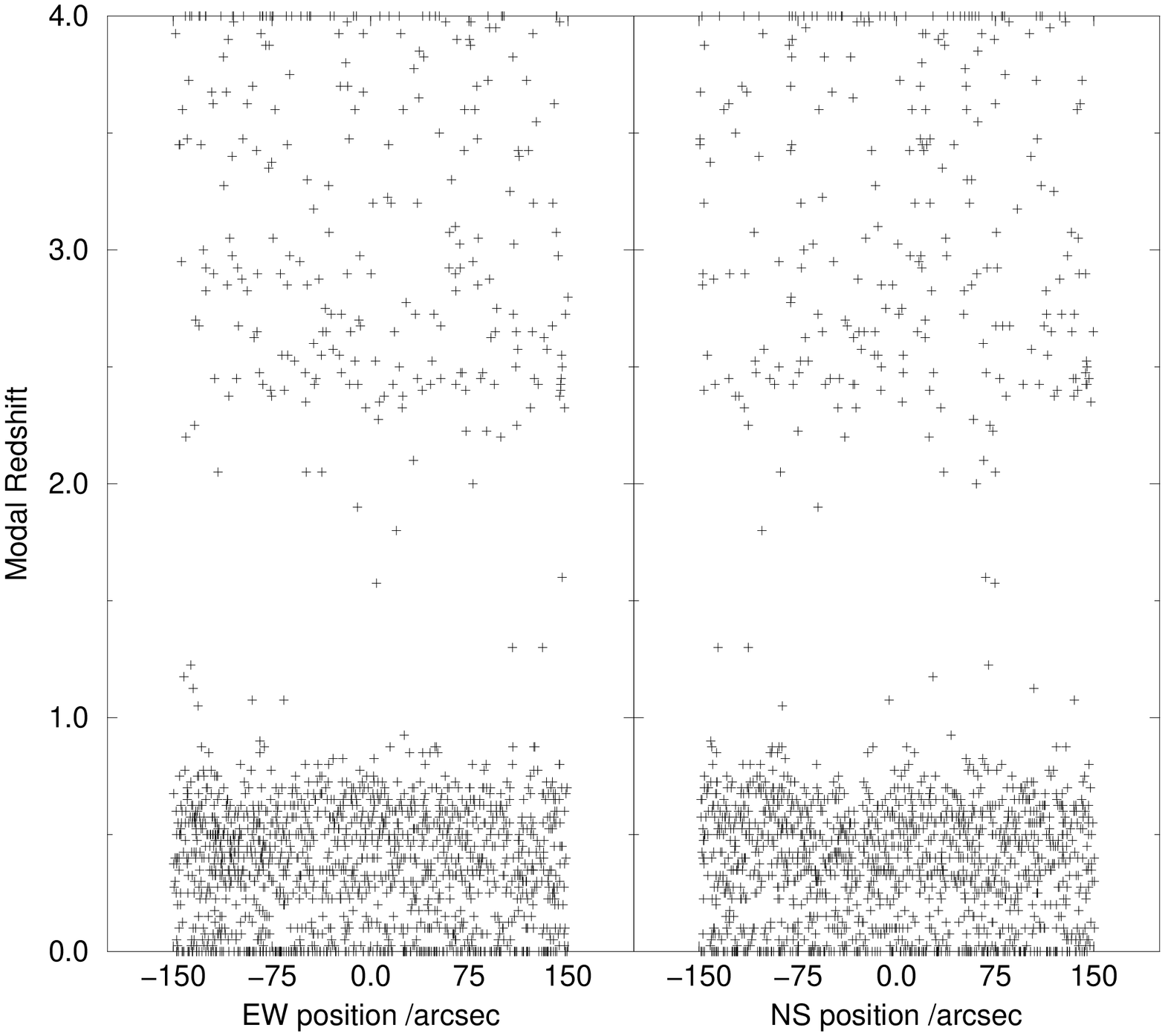,angle=270,width=0.7\linewidth}
\caption{Angular distributions versus redshift for all objects with
$R<26.0$. EW~position is the offset from the centre of the
field along the X--axis of the CCD. NS~position is the offset form the
centre of the position along the Y-axis of the CCD.
\label{fig:zdist}
}
\end{center}
\end{figure}

The two redshift distribution histograms, based on mean and most probable
redshift, are plotted in Figure~\ref{fig:zhist}. The main features of
both these diagrams are the two peaks: one at $0<z<0.6$, and the other
at $z\approx2.4$, with only a small number of objects lying in the
range $1.0<z<2.2$. This is reinforced by studying the distributions of
objects in space against redshift -- these are plotted in
Figure~\ref{fig:zdist}.

It is clear that the apparent lack of objects in this redshift range
is not a real effect -- there cannot really be so few objects in this
region. Identifying the cause of this effect is important for
understanding the limitations of the photometric redshift technique.

\begin{figure}
\begin{center}
\epsfig{figure=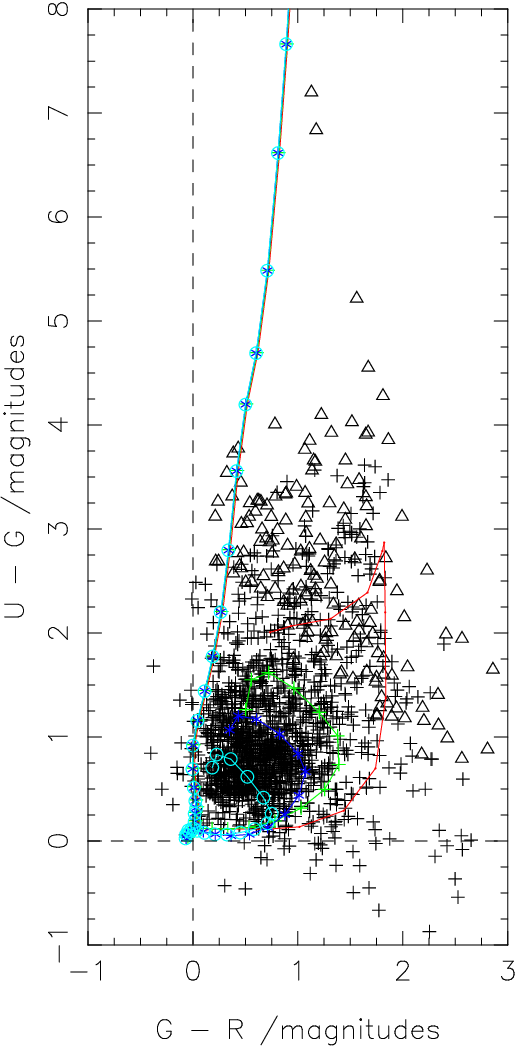,angle=270,width=0.5\linewidth}
\caption{U-G vs G-R plot overlaid with the revised models colour
tracks used in the determination of the photometric redshifts.
\label{fig:uggrcomp}
}
\end{center}
\end{figure}

From the model colour tracks plotted against the actual objects
colours (Figure~\ref{fig:uggrcomp}), it is clear that the
close--packing of the model colours in the redshift range $1.0<z<2.1$
is a problem. This arises because the continuum of the galaxy spectra
is almost flat for all spectral types at wavelengths between 1400\AA\
and 2500\AA . Only the $R-I$ colours provide any redshift sensitivity
for objects with $1.0<z<1.5$. The model colours effectively act as an
'attractor' for objects with similar colours -- 'attracted' objects
are labelled with the redshift and type of the model they end up
with. Thus closely spaced models are less effective at distinguishing
objects, unless there are significantly more objects concentrated in
that area. Even then, the similarity of the objects is sufficient to
make the accurate determination of the redshift impossible in this
redshift range given these filters. The result is that objects with
redshifts $1.0<z<2.1$ end up being pushed to lower ($z\approx0.8$) or
higher ($z\approx2.4$) estimated redshifts.

Beyond the limitations imposed by the filter selection, it is also
notable that the number of objects that are close to the models SED
colours is small in the redshift range $1.0<z<2.0$. This suggests that
the model colours being applied here do not directly correspond to
observed colours in the field. This raises several possibilities:

\begin{enumerate}[(1)]
	\item the model colours are inaccurate;
	\item there are significant systematic errors in the catalogue
	calibration;
	\item the model colours are correct, but the observed colours
	differ due to reddening caused by the presence of dust in
	these objects.
\end{enumerate}

The accuracy of a catalogue's photometric measurements can be
qualified by examining the accuracy of the photometric calibration,
both as determined from the original calibrators and in comparison
with other published results. All the calibrators used in the creation
of the catalogue have high signal--to--noise, and even including the
possibility of poor background level determination, the calibrations
are at worst inaccurate by $\pm 0.05$ magnitudes in all filters. This
is not sufficient to bias the results as dramatically as observed
here. In comparison with other published photometric measurements for
the field, such as Hu \& Ridgway's multi--filter imaging around Quasar
A (\cite{HuR94}), the results are all consistent within the
photometric errors.

\begin{figure}
\begin{center}
\epsfig{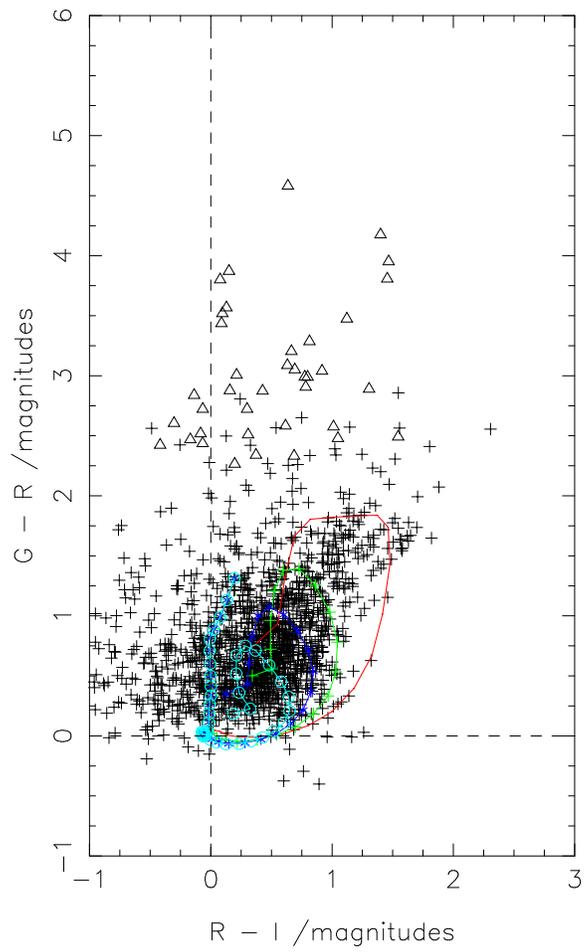}
\caption{$G-R$ vs $R-I$ plot overlaid with the revised models colour
tracks used in the determination of the photometric redshifts.
\label{fig:grricomp}
}
\end{center}
\end{figure}

Since the model colours here are based on unreddened spectra, it seems
reasonable to suggest that the lack of objects with such blue colours
is due to reddening of these objects by dust. This would have the
greatest effect in the $U$ \& $G$ colours, since these are the ones
with the shortest wavelengths. If we examine the spread of colours of
the objects in $G-R$ and $R-I$ (Figure~\ref{fig:grricomp}) we can see
there is better coverage of the model tracks in $R-I$ than in
$G-R$. The objects at $R-I<-0.5$ are all within 2-$\sigma$ of $R-I=0$;
ie this is most probably photometric scatter. This is particularly
noticeable in $R-I$ because of the lower depth of the $I$ image
compared with the rest of the broadband colours.

\subsection{Most likely or mean redshift?}

\begin{figure}
\begin{center}
\epsfig{figure=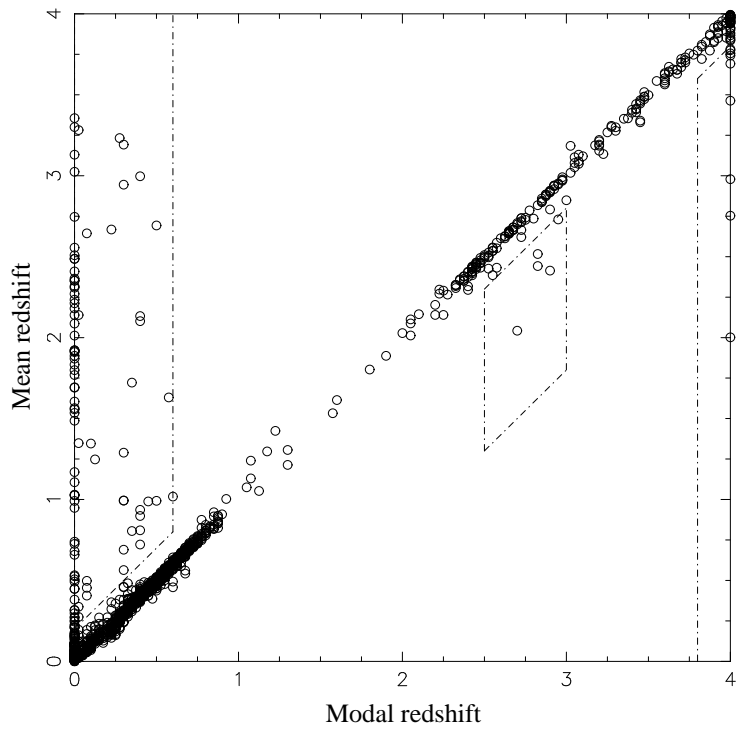,angle=0,width=0.8\linewidth}
\caption{Most likely redshift plotted against the mean redshift
\label{fig:modeormeanz}
}
\end{center}
\end{figure}

Comparing the two estimates shows several interesting features, as
illustrated in Figure~\ref{fig:modeormeanz}. In 1486 out of 1665
objects with $R<26.0$ , there is less than 0.1 difference between the
modal and mean redshift. 63 objects differ by 0.1 -- 0.2 in redshift,
and 116 differ by more than 0.2. For those objects whose redshift
determinations differ by greater than 0.2, these divide into three
major groups: those which have a modal redshift
$0<\textrm{modal~}z<0.6$ but have a large mean redshift, a set with
$2.5<\textrm{modal~}z<3.0$ but a mean redshift of
$1.8<\textrm{mean~}z<2.7$, and a set with $\textrm{modal~}z\approx4.0$
but a lower mean redshift.

\begin{figure}
\begin{center}
\epsfig{figure=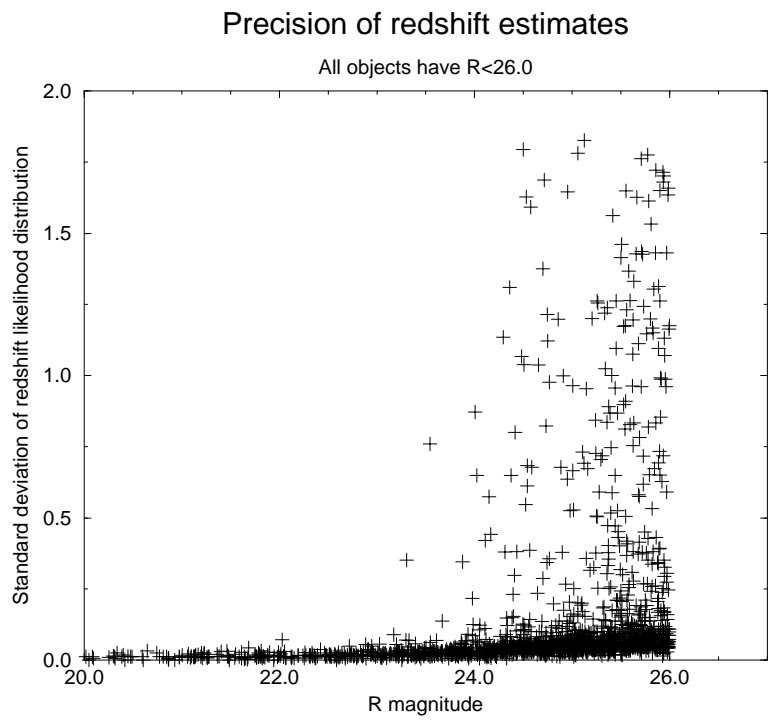,angle=270,width=0.8\linewidth}
\caption{Standard deviations of the redshift distributions for each
object plotted against $R$ magnitude.
\label{fig:r.vs.zdiff}
}
\end{center}
\end{figure}

\begin{figure}
\begin{center}
\epsfig{figure=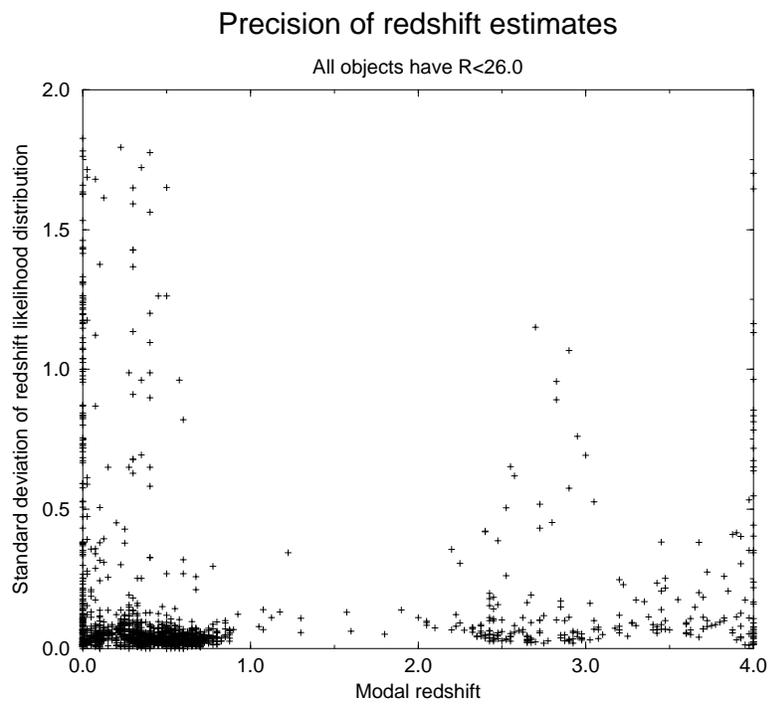,angle=270,width=0.8\linewidth}
\caption{Standard deviations of the redshift distributions for each
object plotted against modal redshift.
\label{fig:modezerr}
}
\end{center}
\end{figure}

In Figure~\ref{fig:r.vs.zdiff}, we plot the precision of the redshift
estimates as a function of $R$ magnitude. The objects which show the
greatest differences are all faint ($R>24.0$); the $\chi^2$ fitting is
looser due to the greater flux uncertainties. This in turn leads to
more models sharing similar model--fitting likelihoods and a greater
spread of redshifts with which the models are statistically
acceptable. Examination of the precision of the redshift estimates
against the modal redshift (Figure~\ref{fig:modezerr}) shows that the
majority of objects with $0<\textrm{modal~}z<0.8$ have small standard
deviations, while the fitting is generally less constrained at higher
redshifts ($z>2.0$) although the average standard deviations even at
$z\approx3$ is still $\Delta z\approx 0.1$.

\begin{figure}
\begin{center}
\epsfig{figure=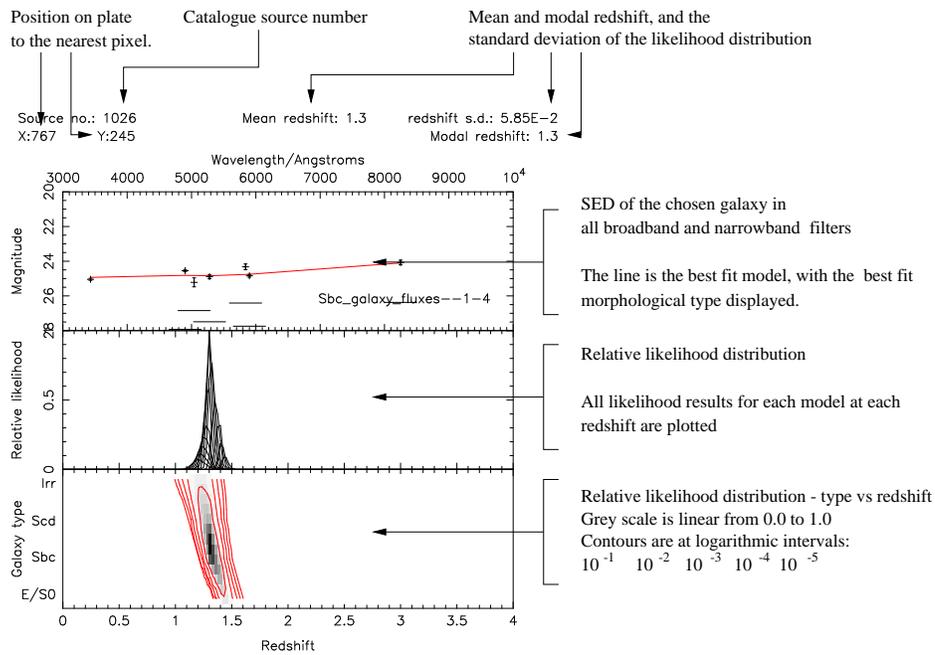,angle=0,width=\linewidth}
\caption{Explanation of the layout of the likelihood distribution graphs
\label{fig:graphdescr}
}
\end{center}
\end{figure}

\begin{figure}
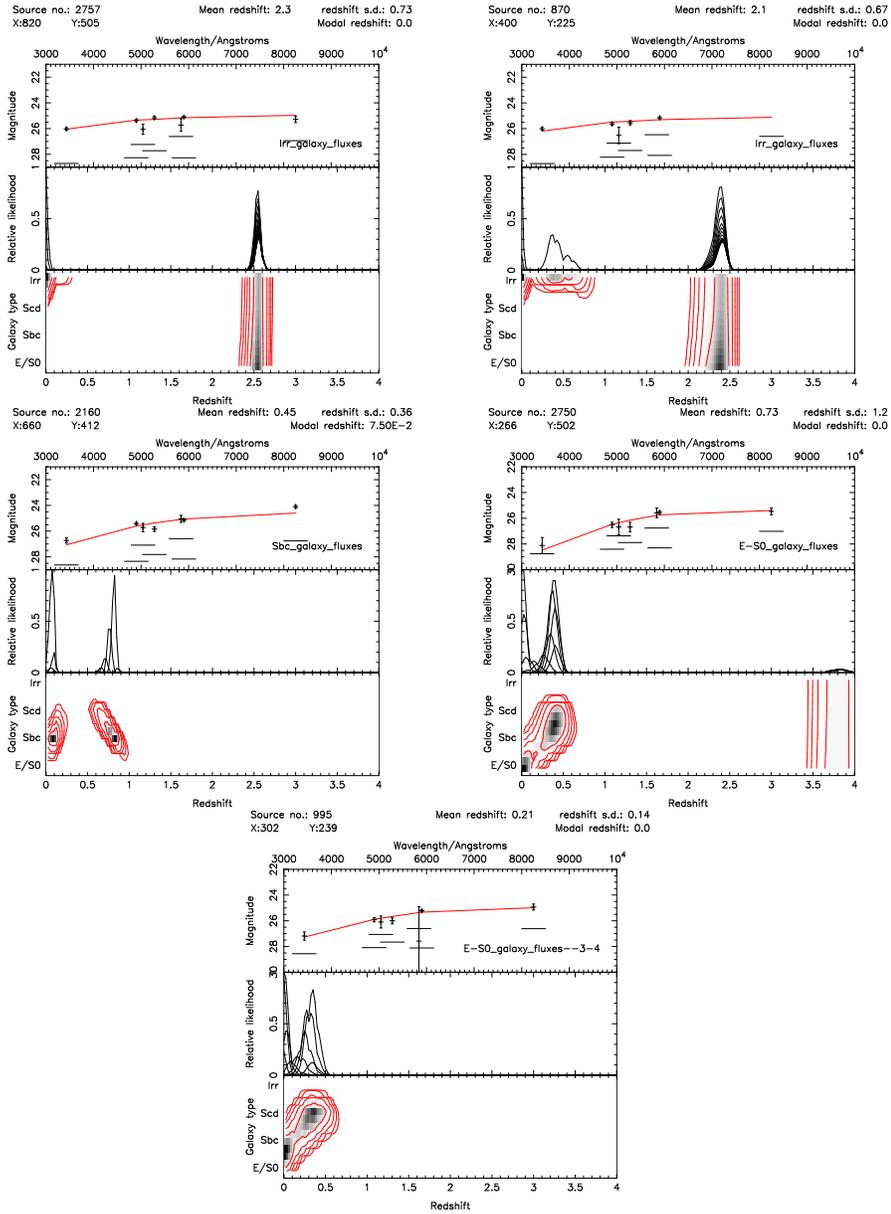

\begin{center}
\begin{minipage}[hp]{0.5\linewidth}
\epsfig{figure=figure16.ps,angle=270,width=0.85\linewidth}
\end{minipage}\hfill
\begin{minipage}[hp]{0.5\linewidth}
\epsfig{figure=figure17.ps ,angle=270,width=0.85\linewidth}
\end{minipage}
\begin{minipage}[hp]{0.5\linewidth}
\epsfig{figure=figure18.ps,angle=270,width=0.85\linewidth}
\end{minipage}\hfill
\begin{minipage}[hp]{0.5\linewidth}
\epsfig{figure=figure19.ps,angle=270,width=0.85\linewidth}
\end{minipage}
\begin{minipage}[hp]{0.5\linewidth}
\epsfig{figure=figure20.ps ,angle=270,width=0.85\linewidth}
\end{minipage}
\caption{Examples of five objects with low modal redshifts but complex
likelihood distributions
\label{fig:lowzconf}
}
\end{center}
\end{figure}

The low--redshift group ($0<\textrm{modal~}z<0.6$) consists of 102
objects, all of which have $24<R<26$. Some examples of the object
spectra and redshift likelihood distributions are shown in
Figure~\ref{fig:lowzconf}, and an explanation of the information on
these graphs precedes in Figure~\ref{fig:graphdescr}

There appear to be two groups of objects which show significant
differences between mean and modal redshift: faint blue galaxies
which, by the nature of their flat and relatively featureless
continuum, are difficult to accurately determine the redshift for
without good photometric constraints; and objects with weak
constraints on the $U$ magnitude which can mimic both $z\approx0$
galaxies and spirals at $z\approx1.0$.

\begin{figure}
\begin{center}
\begin{minipage}[hp]{0.5\linewidth}
\epsfig{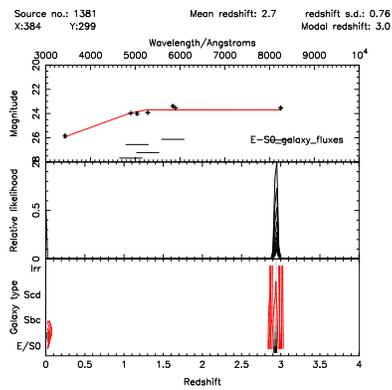}
\end{minipage}\hfill
\begin{minipage}[hp]{0.5\linewidth}
\epsfig{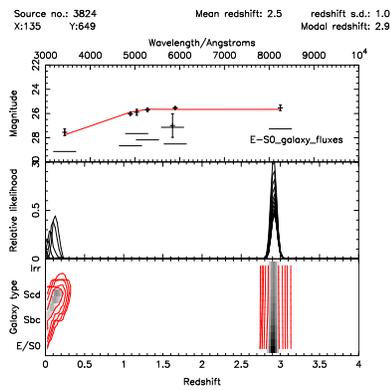}
\end{minipage}
\caption{Two intermediate modal redshift objects with complex
likelihood distributions
\label{fig:medzconf}
}
\end{center}
\end{figure}

\begin{figure}
\begin{center}
\begin{minipage}[hp]{0.5\linewidth}
\epsfig{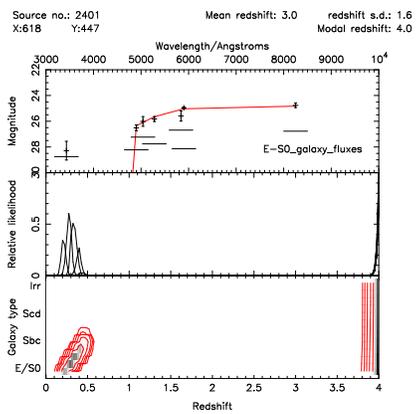}
\end{minipage}\hfill
\begin{minipage}[hp]{0.5\linewidth}
\epsfig{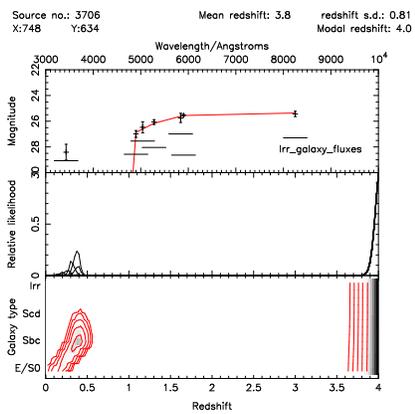}
\end{minipage}
\caption{Two high modal redshift objects with complex likelihood distributions
\label{fig:highzconf}
}
\end{center}
\end{figure}

The intermediate ($2.5<\textrm{modal~}z<3.0$) and high
($3.8<\textrm{modal~}z<4.0$) redshift galaxies also show a fairly flat
spectrum in $G$,$V$,$R$ \& $I$, and red $U-G>$ colours (see
Figures~\ref{fig:medzconf}~\&~\ref{fig:highzconf} for examples). Again
this ambiguity between a low redshift E/S0/Sa galaxy with a strong
4000\AA\ break and a high--redshift starforming galaxy with absorption
due to the Ly$-\alpha$ forest and extinction below 912\AA\ causes
problems for the photometric redshift models. $J$, $H$ and other IR
imaging would break this degeneracy, as would significantly deeper $U$
imaging.

In summary, the modal redshift appears to give the better redshift
determination. When ambiguities arise in the determination of the
redshift, the standard deviation of the redshift is sufficiently
sensitive to mark the problem cases.

\subsection[Low--redshift galaxies]{Low--redshift galaxies -- $z<1.0$}

The accuracy of the photometric redshifts is likely to be best in this
sub--sample, especially in the range $0<z<0.5$. In a blind test
carried out by Hogg et al\nocite{HCB98}, the rms deviation of the
photometric redshift from the spectroscopic determination was
routinely $\Delta z<0.1$, using similar methods and a similar range of
wavelengths covered by the filters used here.

\begin{figure}
\begin{center}
\begin{minipage}[hp]{0.5\linewidth}
\epsfig{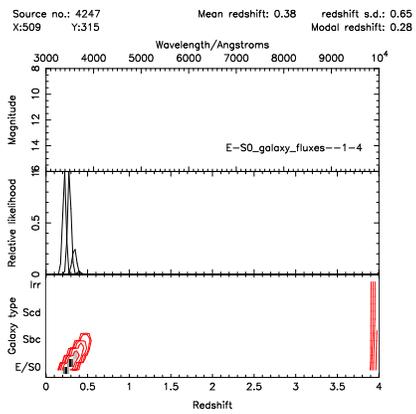}
\end{minipage}\hfill
\begin{minipage}[hp]{0.5\linewidth}
\epsfig{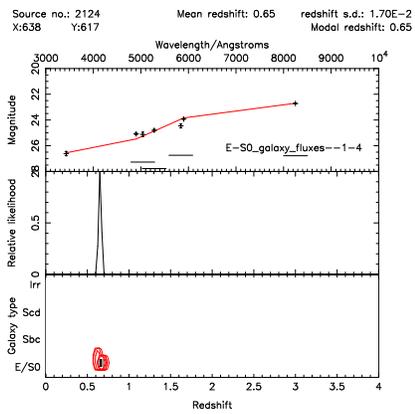}
\end{minipage}
\caption{Two low--redshift E/S0--type galaxies
\label{fig:lowzES0}
}
\end{center}
\end{figure}

\begin{figure}
\begin{center}
\begin{minipage}[hp]{0.5\linewidth}
\epsfig{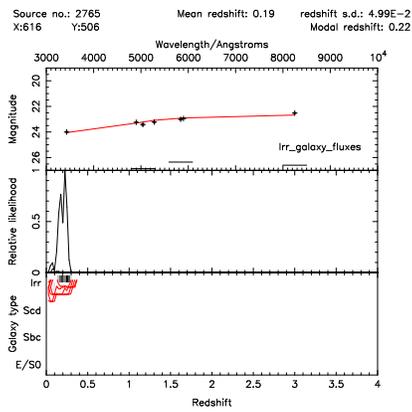}
\end{minipage}\hfill
\begin{minipage}[hp]{0.5\linewidth}
\epsfig{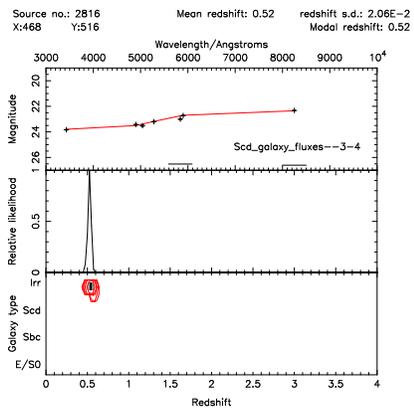}
\end{minipage}
\caption{Two low--redshift Irr--type galaxies
\label{fig:lowzIrr}
}
\end{center}
\end{figure}

Examples of redshift estimations are shown in
Figures~\ref{fig:lowzES0} -- \ref{fig:lowzIrr}. These also include
estimations of galaxy morphological type, made possible by the wide
spread of colour space covered by the photometric models at low
redshift.

A few spectroscopic redshifts for some of the galaxies in this field
are already available, obtained by one of us with the WHT allowing one
to check the accuracy of the photometric
predictions. Table~\ref{tab:compsp} shows that the photometric
redshifts are accurate to $\Delta z\approx0.15$, although the
systematic offset suggests that the models can be further improved. A
similar overestimate is seen in \cite{LYF96} where the photometric
models are based on redshifted CWW spectra alone, where the rms
scatter was reported as $\Delta z\approx0.15$. Comparing the published
redshift for HR10 (\cite{GD96}), it is clear that the reddening of the
spectrum in this $z>1$ galaxy has resulted in an erroneous
estimate.

\subsection{Intermediate--redshift galaxies -- $1.0<z<2.5$}

\label{sec:intz}

As discussed in section~\ref{sec:over} there are too few objects
identified in the range $1.0<z<2.5$. It is important to determine what
redshifts objects in this redshift range have in fact been
given. Assuming that the models are at least a reasonable pointer to
the correct colours, possibly affected by reddening, objects at these
redshifts will have been mis-classified towards those models which
have colours similar to the $1.0<z<2.1$ models. The most likely cases
are therefore the low--redshift Irregular galaxies, which have
relatively flat spectra in the visible wavelengths, and those models
at redshifts $z\approx 0.8$ and $z\approx2.3$ which are adjacent
region we are interested in. This can actually be seen in the
histogram of objects against modal redshifts (Figure~\ref{fig:zhist})
-- there is an apparent excess of low--redshift objects at $z=0$ along
with another peak in the distribution around $z=2.4$. No excess is
seen at $z\approx0.8$ but any increase here would be likely to be a
small fraction of the real objects at this redshift.

\begin{figure}
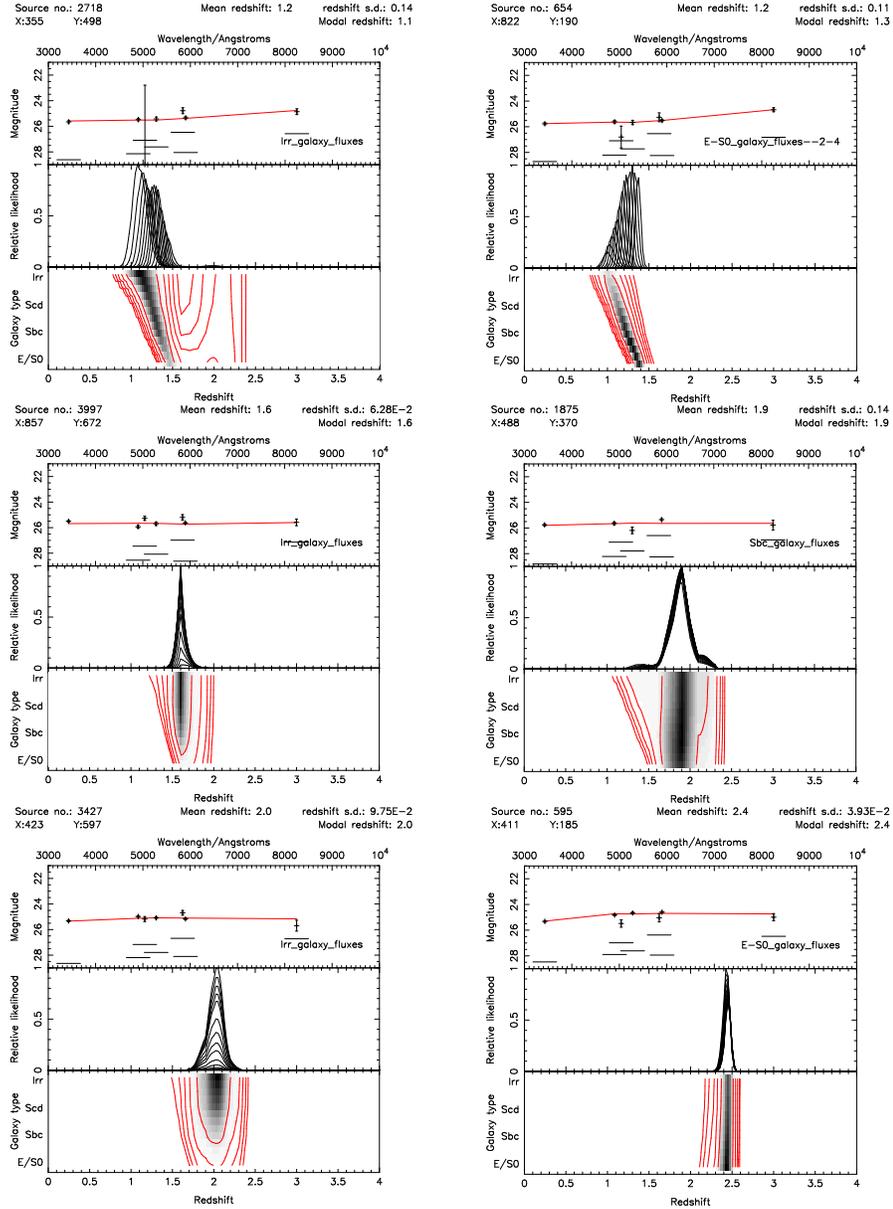

\begin{center}
\begin{minipage}[hp]{0.5\linewidth}
\epsfig{figure=figure33.ps,angle=270,width=0.85\linewidth}
\end{minipage}\hfill
\begin{minipage}[hp]{0.5\linewidth}
\epsfig{figure=figure34.ps,angle=270,width=0.85\linewidth}
\end{minipage}
\begin{minipage}[hp]{0.5\linewidth}
\epsfig{figure=figure35.ps,angle=270,width=0.85\linewidth}
\end{minipage}\hfill
\begin{minipage}[hp]{0.5\linewidth}
\epsfig{figure=figure36.ps,angle=270,width=0.85\linewidth}
\end{minipage}
\begin{minipage}[hp]{0.5\linewidth}
\epsfig{figure=figure37.ps,angle=270,width=0.85\linewidth}
\end{minipage}\hfill
\begin{minipage}[hp]{0.5\linewidth}
\epsfig{figure=figure38.ps,angle=270,width=0.85\linewidth}
\end{minipage}
\caption[Six examples of galaxies with fits to model spectra $1.0<z<2.5$]
{Six examples of galaxies with fits to model spectra $1.0<z<2.5$. Note
that there is little constraint on the morphology of the objects shown
here. Additionally, at $z\approx1.3$ there is a degeneracy between
morphology and redshift.
\label{fig:intzspectra}
}
\end{center}
\end{figure}

Those objects which are identified at $1.0<z<2.5$ -- examples are shown
in Figure~\ref{fig:intzspectra} -- are only weakly distinguished in
morphological type. Additionally there is a degeneracy between
morphological type and redshift estimate for galaxies with redshifts
$1.0<z<1.6$ -- at higher redshifts there are few major differences
between the galaxy colour tracks and hence no significant
discrimination between morphology based on colour evidence. To remove
these degeneracies would require more IR imaging to detect the
4000\AA\ break -- this would also improve the accuracy of the redshift
detections in this redshift range.

\subsection{High--redshift galaxies -- $2.5<z<4.0$}

In Paper I, we have tried to identify galaxies at $z\approx3$ via the
912\AA\ break. Photometric redshifts should prove effective at
determining which galaxies lie at these redshifts since they make use
of all the photometric measurements together in all filters, rather
than just the $U$,$G$ \&\ $R$. 

\begin{figure}
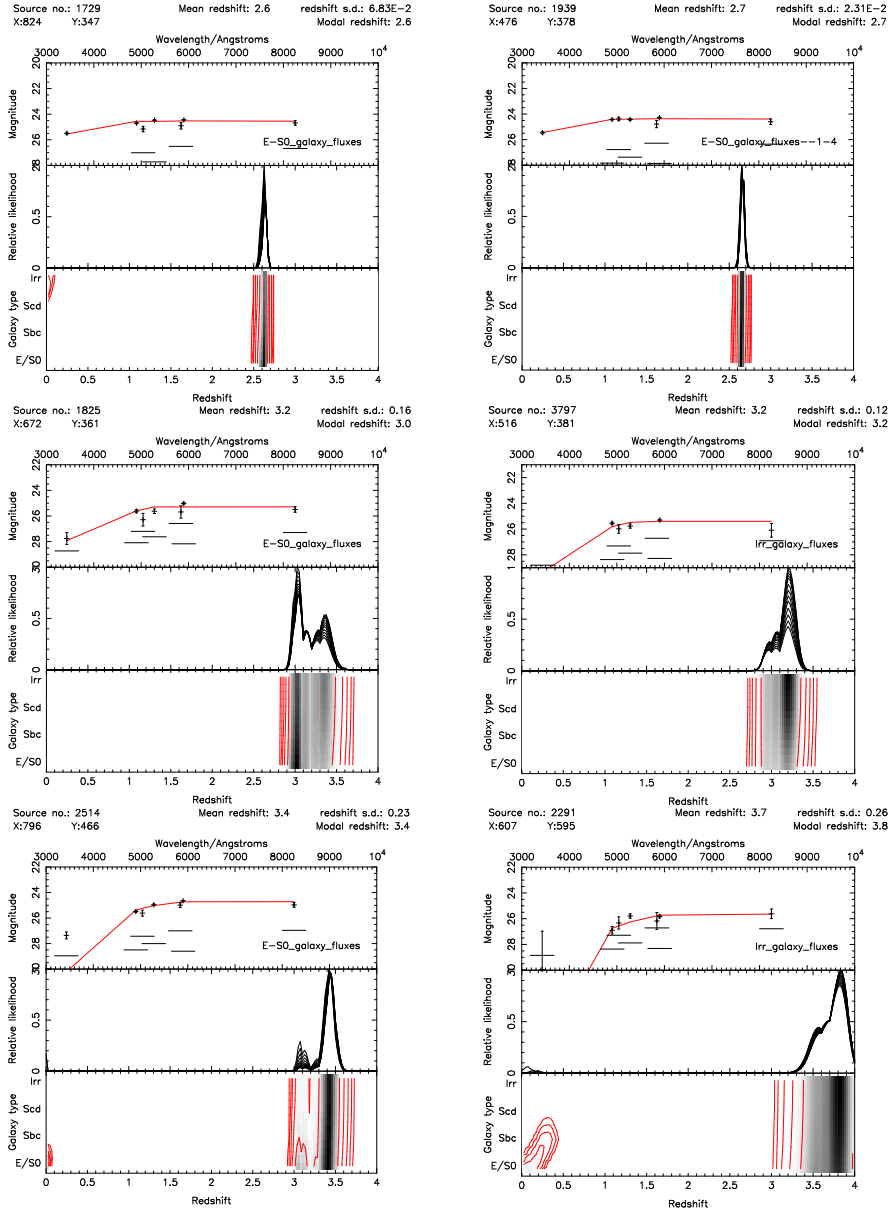

\begin{center}
\begin{minipage}[hp]{0.5\linewidth}
\epsfig{figure=figure39.ps,angle=270,width=0.85\linewidth}
\end{minipage}\hfill
\begin{minipage}[hp]{0.5\linewidth}
\epsfig{figure=figure40.ps,angle=270,width=0.85\linewidth}
\end{minipage}
\begin{minipage}[hp]{0.5\linewidth}
\epsfig{figure=figure41.ps,angle=270,width=0.85\linewidth}
\end{minipage}\hfill
\begin{minipage}[hp]{0.5\linewidth}
\epsfig{figure=figure42.ps,angle=270,width=0.85\linewidth}
\end{minipage}
\begin{minipage}[hp]{0.5\linewidth}
\epsfig{figure=figure43.ps,angle=270,width=0.85\linewidth}
\end{minipage}\hfill
\begin{minipage}[hp]{0.5\linewidth}
\epsfig{figure=figure44.ps,angle=270,width=0.85\linewidth}
\end{minipage}
\caption{Examples of galaxies with fits to model spectra
$z>2.5$. There is no possibility of morphological classification
at these redshifts. Note that at $z\gtrsim3.4$ the effect of the
912\AA\ break moving into the blue end of the $G$ band starts to
redden the $G-R$ colours, and that this feature is identified by the
photometric redshift technique more strongly than the faint $U$
magnitudes, which have large photometric uncertainties.
\label{fig:highzspectra}
}
\end{center}
\end{figure}

Examples are shown in Figure~\ref{fig:highzspectra}. It is not possible to
differentiate between different morphological types at these redshifts
-- the colours are too similar between the model galaxies. This is
partly a result of similar star--formation histories between the model
galaxies at these redshifts; this would be invalid if, for example,
there were several phases of galaxy formation.

\begin{figure}
\begin{center}
\epsfig{figure=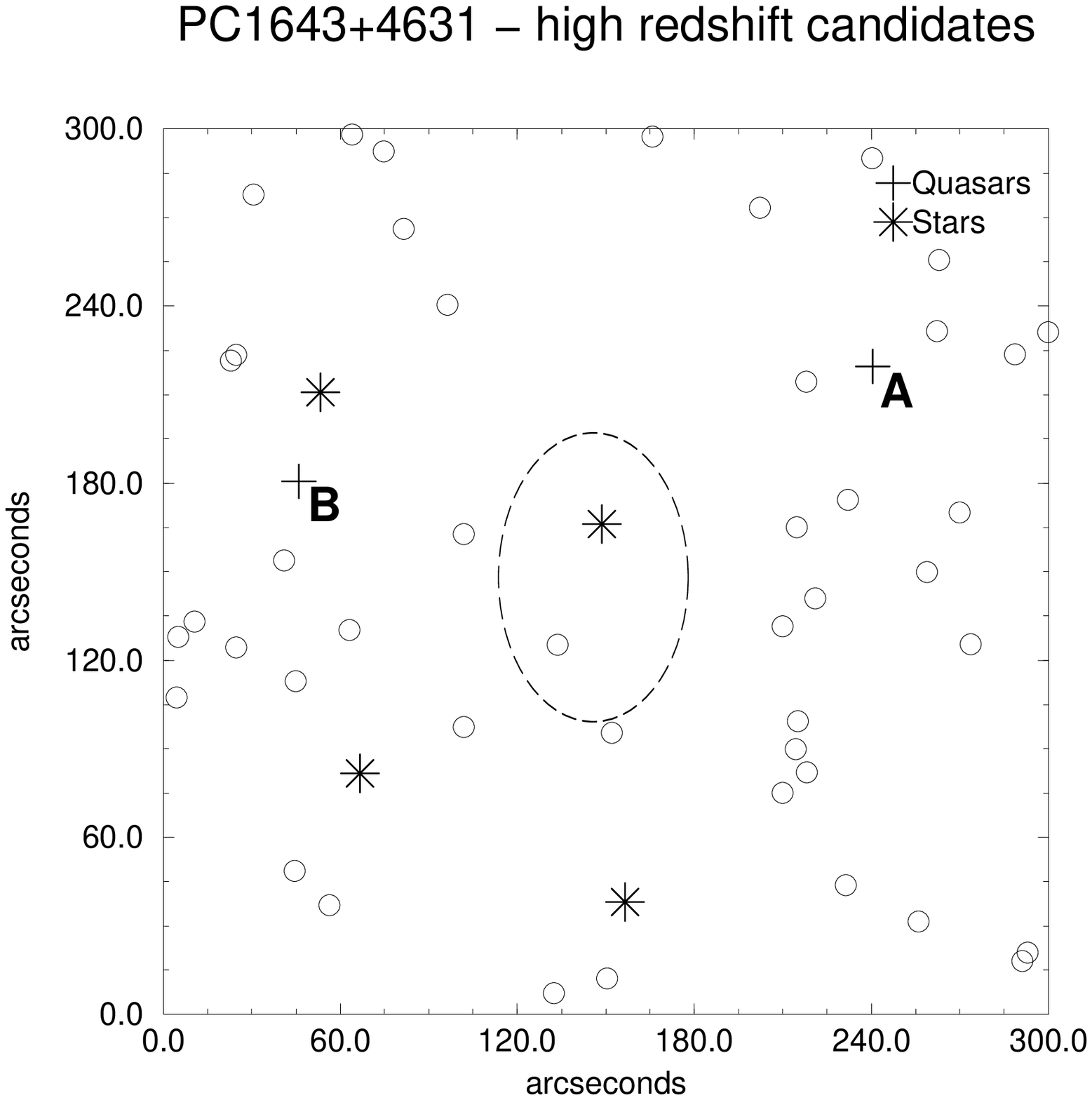,angle=270,width=0.8\linewidth}
\caption{The distribution of the Ly--break candidates, which are
marked as circles, selected using $R<25.5$, $G>2\sigma$ and
$2.8<\textrm{modal~}z<3.5$.
\label{fig:zapprox3photz}
}
\end{center}
\end{figure}

By selecting galaxies with $2.8<\textrm{modal~}z<3.5$, $R<25.5$ and
$G>2\sigma$, we get 46 Ly--break candidates, almost twice as many as
with the colour--criteria. The distribution of these galaxies (see
Figure~\ref{fig:zapprox3photz}) also (cf Paper I) shows an apparent
hole at a point mid--way between the quasars, despite the increase in
surface density of candidates.

\section{Clustering of objects}

\begin{figure}
\begin{center}
\epsfig{figure=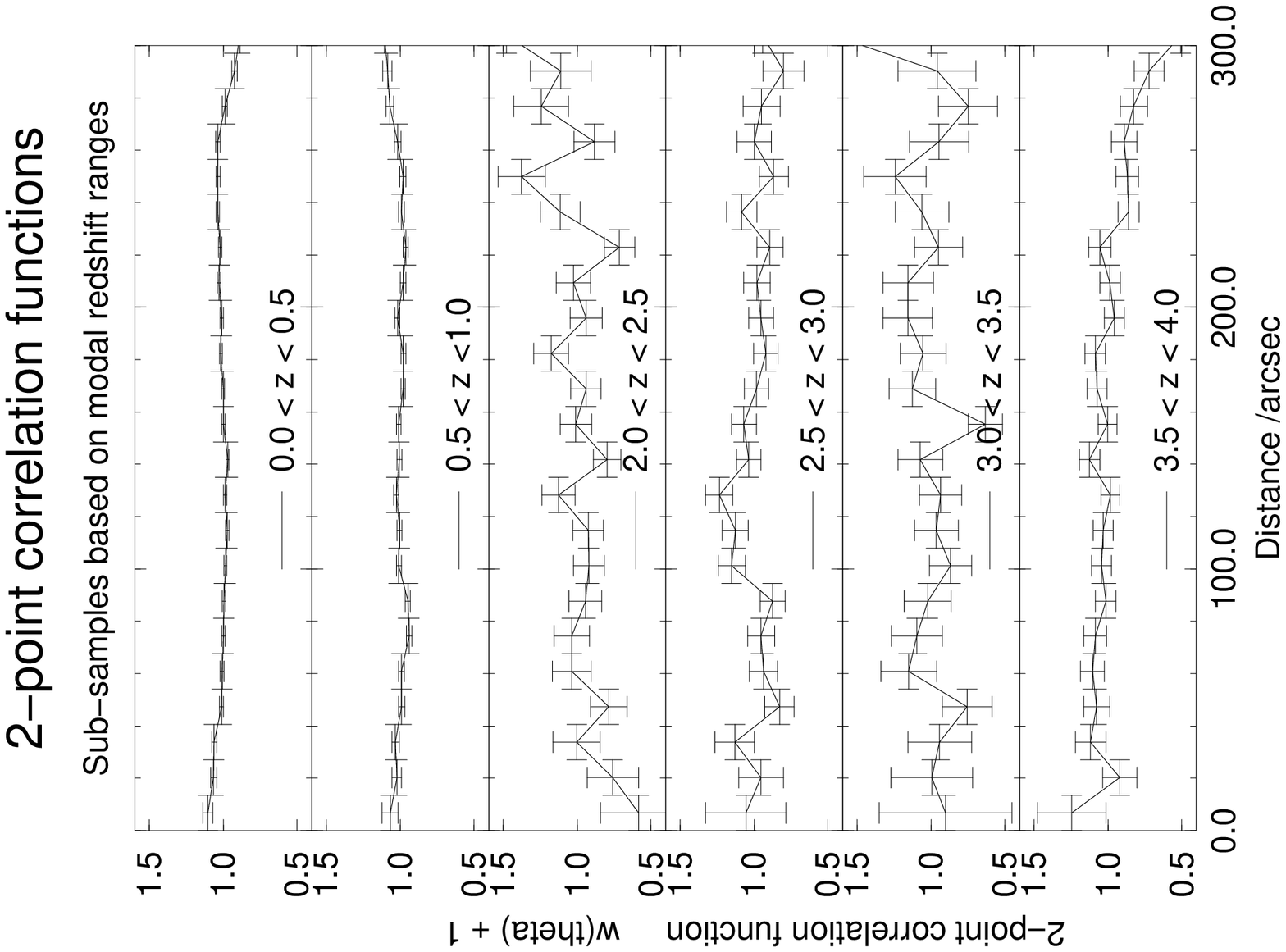,angle=270,width=0.8\linewidth}
\caption{Comparison of the clustering of the galaxies in the field in
different redshift ranges
\label{fig:compcorrel}
}
\end{center}
\end{figure}

We can make use of the redshift determinations to look at the
correlation functions for redshift subsets of the catalogue. The
results of these are plotted in Figure~\ref{fig:compcorrel}.

\begin{figure}
\begin{center}
\epsfig{figure=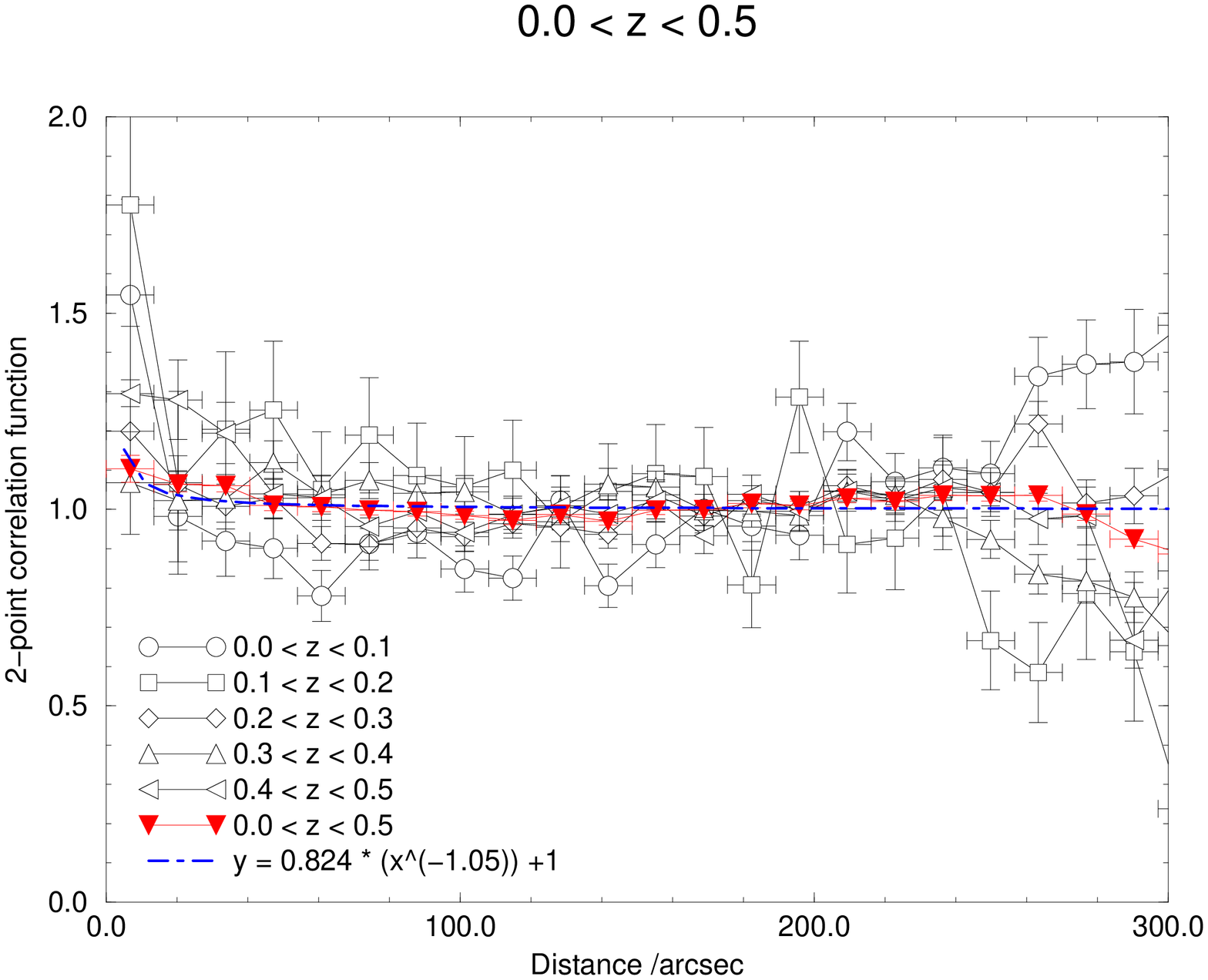,angle=270,width=0.8\linewidth}
\caption{Correlation function for galaxies with modal redshifts $0<z<0.5$
\label{fig:0to0.5correl}
}
\end{center}
\end{figure}

\begin{figure}
\begin{center}
\epsfig{figure=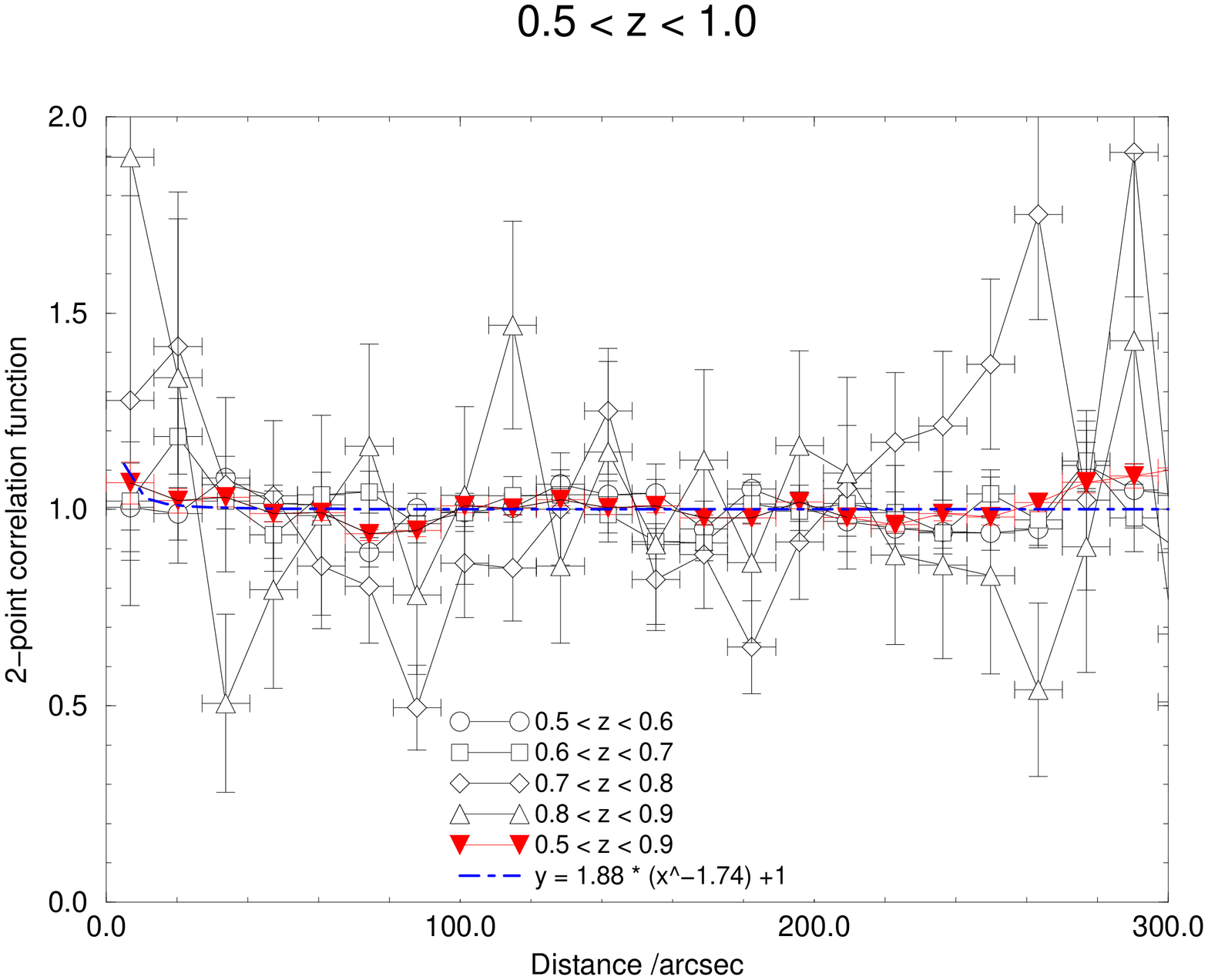,angle=270,width=0.8\linewidth}
\caption{Correlation function for galaxies with modal redshifts $0.5<z<1.0$
\label{fig:0.5to1correl}
}
\end{center}
\end{figure}

The lowest redshift range ($0<z<0.5$) has 558 objects, and there are
352 objects in the range $0.5<z<1.0$. As a result, the statistics for
these redshift ranges are sufficiently good to show that there is
measurable clustering in this field at low redshifts. Fitting the
standard power--law curves ($A_\omega\theta^{-\beta}$) to angular
scales $0<\theta<150~\textrm{arcsec}$ gives $A_\omega\approx0.82$ and
$\beta\approx1.05$ for $0.0<z<0.5$, and $A_\omega\approx1.88$ and
$\beta\approx1.74$ for $0.5<z<1.0$. These are plotted in
Figures~\ref{fig:0to0.5correl} \& \ref{fig:0.5to1correl}. Also plotted
in these figures are the correlation functions for subsets of the
sample ordered in redshift. While the noise in these subsets is worse
and no strong conclusions can be drawn, the apparent clustering of the
$0.7<z<0.9$ objects at small angular scales is interesting.

In the rest of the $z$ bins there is little significant signal, with
the least insignificant being 1.5-$\sigma$ clustering at $240''$ in
$2.0<z<2.5$. Necessarily, there is no coverage of $1.0<z<2.0$.

\section{Artificial clusters}
We now test the ability of this sort of analysis to recover real
features from the field. We use these redshift colour models to
place a realistic spatial distribution of artificial galaxies into the
images and examine the likelihood of detecting a cluster towards
PC1643+4631 A\&B using colour methods. We can find no similar
experiment in the published literature on any field.

\subsection{Theoretical magnitude and spatial distributions}

To mimic a cluster of galaxies, one needs distributions for both the
magnitudes and positions of the galaxies on the sky. For the
former we assummed a Schechter luminosity function
(\cite{Schechter}), while for the latter we assumed a Hubble
distribution, where the surface density is described by
$\sigma_0/(1+(R/R_c)^2)$ (\cite{WH93}). We used a core radius,
$R_c=200$kpc for the spatial distribution (Jones et al., 1997, found,
for PC1643, a core radius of $\sim300$kpc), and used half--light radii of
10kpc for the galaxies.

\sloppy{To determine the magnitudes of galaxies in a cluster at some redshift,
we used the results from the Canada France Redshift Survey (CFRS)
(\cite{CFRS1}), because it is one of the best measured in a constant
rest frame waveband and it provides us with the best--fit $B$--band
Schechter luminosity function for $z\gtrsim0.5$ galaxies
(\cite{LTH95}). The CFRS contains many objects at redshifts out to
$z\approx1.1$, and provides the best estimate of the luminosity at
$z\approx0.5$. Additionally, for $z<0.9$, the $B$--band is redshifted
to wavelengths covered by the $U$,$G$,$V$,$R$ and $I$ filters. Taking
the values $\alpha=1.03$, absolute magnitude $M^*_{AB}(B)=-21.04$, we determine the
observed total magnitudes from these galaxies using (see \cite{LTH95})}

\begin{equation}
I_{\textrm{AB}} = M_{\textrm{AB}} + 5 \log (D_L/10pc) - 2.5 \log (1+z)
- (B-I_z),
\end{equation}

\noindent $D_L$ is the luminosity distance, $I_z$ is the $I$ band shifted to
shorter wavelengths by a factor of $(1+z)$, and $B$ is the magnitude
in the rest--frame $B$ band. The value of $(B-I_z)$ is taken from the
photometric SED models, and hence the value of $B$ must be calculated
by interpolation between the $U$, $G$ and other broadband
filters. From this, observed magnitudes can be calculated for the
redshift range $0<z<1.5$. For $z>1.5$, $B$ moves to longer wavelengths
than $I$. A simple extrapolation of the $R-I$ colour can be used to
obtain a first--order estimate of the rest frame $B$ band
magnitude. This should prove to be a reasonable approximation of the
real value given that the 4000\AA\ break is weak in galaxies which are
strongly star--forming, which is the case for the SED models used at
$z>1$.

We do not attempt to model the evolution of the luminosity
function. While this is a shortcoming of this method, at $z<0.9$, the
maximum likely error is $\pm0.5$ mag (see \cite{LTH95}). Beyond
$z\approx1$, the evolution of the luminosity function is poorly
constrained, but it is reassuring that the luminosity function of
$z\approx3$ galaxies have a luminosity function best fitted by
$M_*\approx-21$ (\cite{MD97}).

At low redshifts, there are significant colour differences between
different morphological galaxy types, with spirals having much bluer
colours than ellipticals. While the ratio of ellipticals to spiral
galaxies in general is about 60:40, in the cluster environment the
proportion of elliptical galaxies is often much lower, closer to 40:60
(E+S0:S+Irr) (\cite{DOC97}). To simulate this spread of colours, we
opted to create frames with just one morphological type in each. To
keep the processing as simple as possible two plates were made, with
one having E/S0 type galaxies and the other Scd/Irr types.

\subsection{Method}

\begin{figure}
\begin{center}
\epsfig{figure=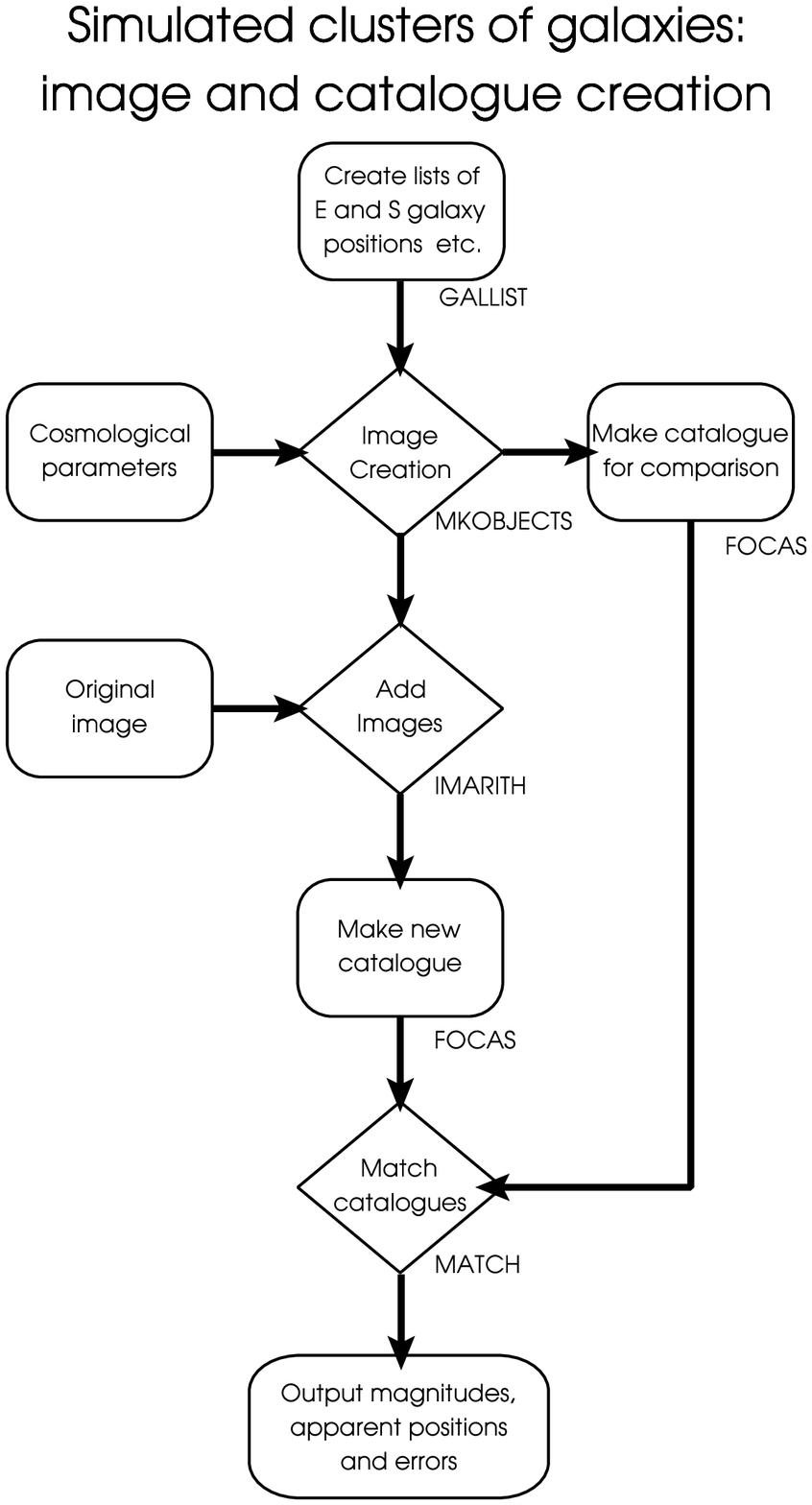,angle=0,width=1.0\linewidth}
\caption{Flow diagram showing the procedures used for simulating
images of clusters of galaxies and obtaining a catalogue from those
images
\label{fig:simclus}
}
\end{center}
\end{figure}

The processes required to simulate each cluster at each redshift and
make photometric measurements of the simulated cluster galaxies after
superimposition onto the PC1643 are illustrated in
Figure~\ref{fig:simclus}. Simulated clusters were made at redshifts
from 0.1 to 4.0 in steps of 0.1.

We used the ARTGAL package in IRAF to create artificial clusters of
galaxies with the required properties. For each `cluster' only two
images were made, one each for elliptical and spiral galaxies, made to
have the same zero point and exposure time as the $I$ image, and zero
background level. No additional noise was simulated because the real
images are all background limited.

Recent publications discussing the distributions of the different
morphological types suggest that there is some segregation, with
steeper velocity dispersions for later--type galaxies (see, eg
\cite{ABM98}). We have not attempted to mimic these results here,
since our aim is to examine the ability of the tests previously
carried out on this field to recover cluster members, rather than to
produce a completely accurate cluster distribution.

Having generated the $I$--band spiral image at each redshift, we used
the Sbc model SEDs to determine the colours of the simulated spirals
in $R$,$V$,$G$ \& $U$, and IMARITH was used to scale the $I$ image to
mimic the correct colours in the other broadband images. We used a
similar procedure to produce $R$,$V$,$G$ \& $U$ images from the
$I$--band elliptical image using the E/S0 model SEDs. These $I$, $R$,
$V$, $G$ \& $U$ spiral and elliptical images were summed together to
create composite images, and a catalogue was made from the $R$
composite image using FOCAS, which we will refer to as the matching
composite catalogue (MCC).

These composite images were then added onto the real PC1643
images. The resulting $R$ image was processed using FOCAS to produce a
catalogue based on $R$ isophotal apertures: catalogues were then made
in $I$, $V$, $G$ \& $U$ using the $R$ isophotes and the five resulting
catalogues were then matched using MATCH against the MCC. To extract
the simulated galaxies from these newly created FOCAS catalogues,
`stripped' catalogues were made including objects only if they
corresponded to the positions indicated by the MCC - ie these
catalogues should contain only the information on the measureable
simulated cluster galaxy members. These stripped catalogues were then
given the same photometric redshift analysis as the real objects in
the PC1643 field.

\subsection{Results}

Full--colour images showing the artificial cluster at various
redshifts are shown in
Figures~\ref{fig:0.2-0.8.images}~\&~\ref{fig:1.0-4.0.images}, both in
isolation and superimposed on the PC1643 field. While it is easy to
recognise the cluster on its own, it becomes impossible to
differentiate the cluster against the other objects in the field at
$z>1.0$, and the contrast is poor even at $z=0.6$.

More importantly, it is the ability of the catalogue creation and
analysis software to detect a cluster which is of greatest interest
here, particularly at faint magnitudes $R\gtrsim24$. We have plotted
the $U-G$ vs $G-R$ colours for each of the simulation images
illustrated in Figures~\ref{fig:0.2-0.8.images} \&
\ref{fig:1.0-4.0.images}. Even at $z=0.2$ the fainter members of the
simulated galaxies show significant photometric scatter away from the
expected colours, and this is reflected in the estimates of modal
redshift: there is a peak at $z=0.2$ as expected, but over a third of
the sample are more than 0.1 away from this redshift, and 9 of the
simulated galaxies are mistaken for high--redshift galaxies at
$z\gtrsim3$.

\begin{figure}
\begin{center}
\epsfig{figure=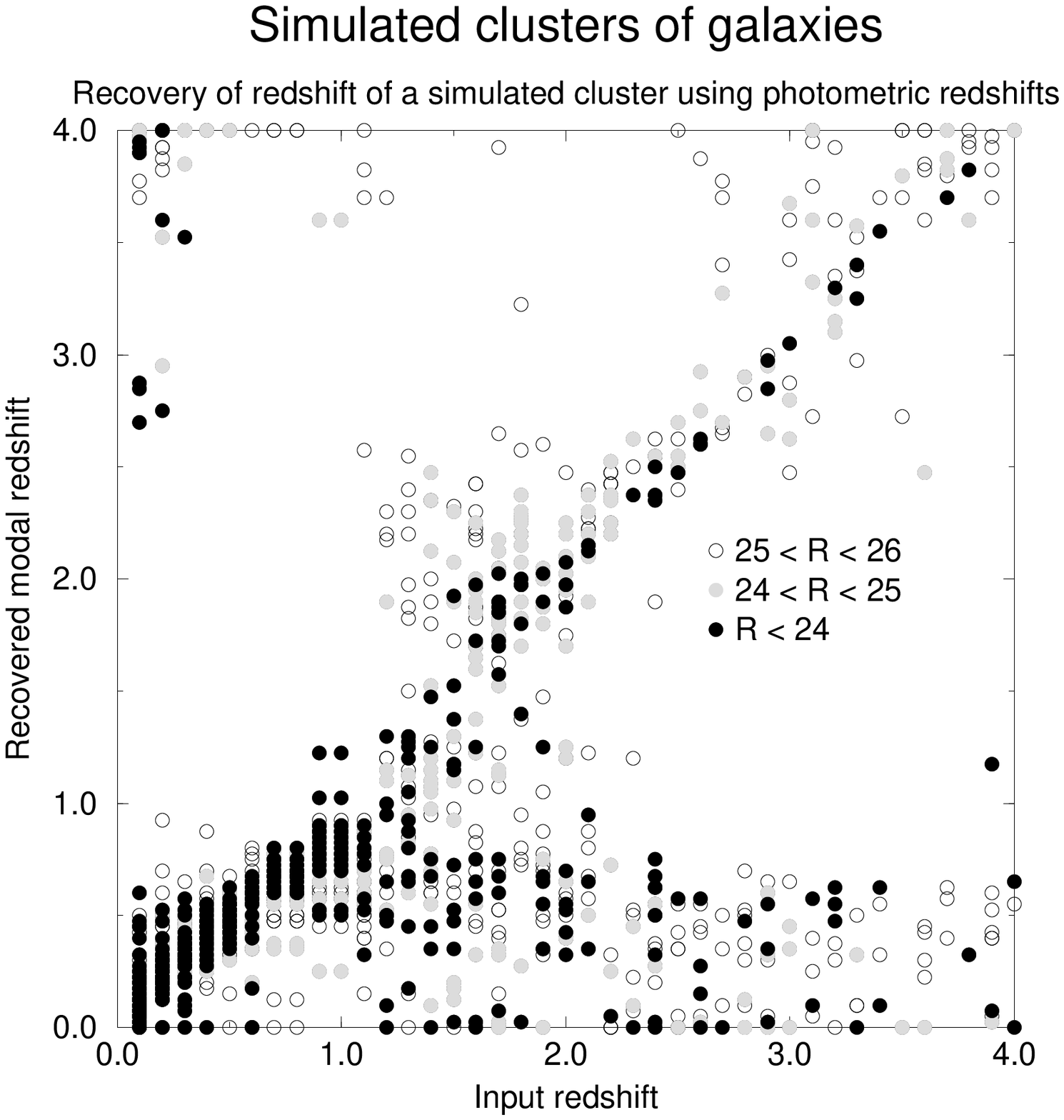,angle=270,width=0.8\linewidth}
\caption{Modal redshift estimates based on photometry from recovering
the member galaxies of simulated clusters at $z=0.1$ to $z=4.0$ in
steps of 0.1 in $z$ using FOCAS. Modal redshifts are shown for all
simulated galaxies which have a measured $R<26.0$.
\label{fig:modalzexp}
}
\end{center}
\end{figure}

\begin{figure}
\begin{center}
\epsfig{figure=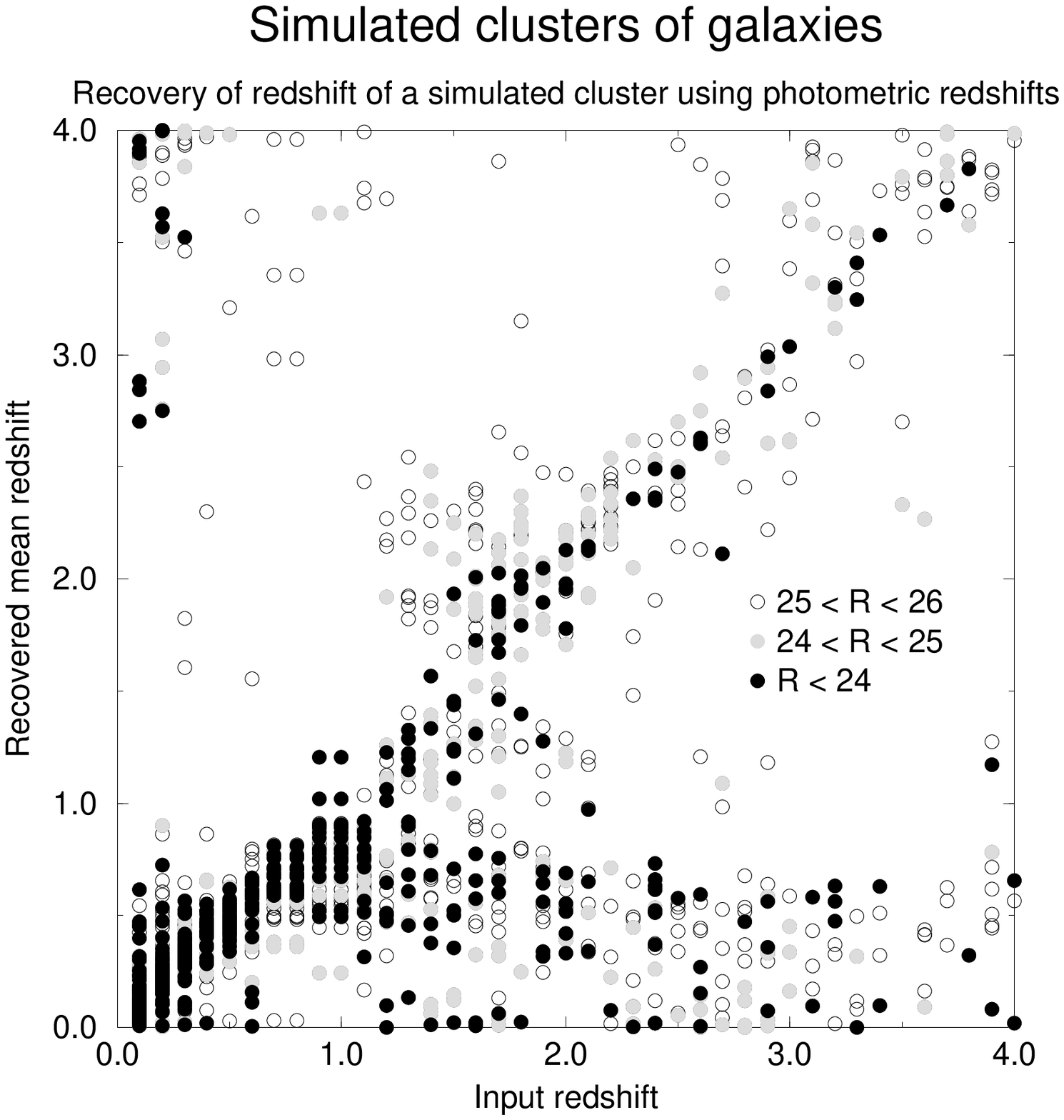,angle=270,width=0.8\linewidth}
\caption{Mean redshift estimates based on photometry from recovering
the member galaxies of simulated clusters at $z=0.1$ to $z=4.0$ in
steps of 0.1 in $z$ using FOCAS. Mean redshifts are shown for all
simulated galaxies which have a measured $R<26.0$.
\label{fig:meanzexp}
}
\end{center}
\end{figure}

\
At $z\lesssim1.0$, the photometric--redshift technique still produces
redshift estimates around the expected value, with increased scatter
as the simulated cluster becomes fainter with increasing redshift (see
Figures~\ref{fig:modalzexp}~\&~\ref{fig:meanzexp}). This is also
reflected in the increased scatter of the $U-G$ and $G-R$ colours,
which, it should be stressed, is often greater than 3-$\sigma$, where
$\sigma$ is the photometric error based on the noise for the
object. This is the result of pollution of the isophotal apertures by
nearby neighbours, which is inevitable in a deep, ground--based field
like this one. Given that the density of objects in the field
including all detections down to the surface brightness limits of
3-$\sigma$ in an area of six pixels is approximately one object in
every $5''\times5''$ box, and that the average isophotal area of a
object with $R=25.0$ is some 5 arcmin$^2$, this means that at least
one in five faint objects will overlap another faint object, so that
in this set of deep images, objects \emph{are being confused}.

To determine whether a significantly brighter cluster of galaxies
would be visible in this field, we repeated the simulation with all
the magnitudes brightened by one magnitude. Bearing in mind that this
is the most extreme error likely in the simulation, this is also a
good test of the sensitivity of these simulations to the real
magnitude distribution. Even with many more members, such a cluster
would still be difficult to distinguish using colour--criteria or by
excess surface density.

At intermediate redshifts, $1.0<z<2.0$, the photometric redshift
technique fails to correctly determine the redshift in almost all
cases. Despite starting from simulated galaxies with the correct
colours, the colours of the galaxies as recovered from the field are
systematically scattered to redder colours. There may be a systematic
effect resulting from selecting objects based on their $R$ magnitudes
--- for example, objects which are polluted to fainter magnitudes in
$R$ are missed from the sample, whereas objects which recover fainter
$G$ magnitudes than originally simulated appear redder than they
really are. There may also be a systematic effect from the different
surface densities of objects in $R$ and $G$ at similar magnitudes --
there are 763 objects with $25<R<26$, and only 537 with $25<G<26$ --
so objects with $G-R\approx0$ stand proportionally more chance of
being brightened in $R$ than in $G$. Additionally, if the position of
a simulated galaxy does coincide with that of a real object, it is
likely that the real object has $G-R\gtrsim0.5$, and hence the
pollution of the simulated object will result in redder measured
colours.

This failure to accurately detect galaxies in the range $1.0<z<2.0$ is
a function of the filters used. Because there are no strong continuum
features over the wavelengths spanned by the optical filters used here
at these redshifts (approximately 1200\AA\ -- 4000\AA) the photometric
redshift are poorly determined. Additionally, because of the scatter
of the objects colours away from the simulated colours, objects with
redshifts in the range $1.0<z<2.0$ are scattered to lower and higher
redshifts. This is clearly seen in
Figures~\ref{fig:modalzexp}~\&~\ref{fig:meanzexp}. The result is that
any histogram of galaxies over the redshifts estimated from
photometric techniques using a similar range of filters to those used
here will show a deficit of galaxies with $1.0<z<2.0$ and peaks above
and below this range against the true population of
galaxies. Precisely this behaviour is seen in the redshift
distribution calculated from the HDF four--colour photometry (in
F300W, F450W, F606W and F814W) using photometric redshifts (see
\cite{GH96} and to a lesser extent \cite{SLY97}). The deficit is not
quite so marked as it is in the photometric redshifts estimates of the
field of PC1643, due to slight differences in modelling but the effect
is still the same. It is notable that the photometric redshift
estimates of \cite{LYF96} (LYFS) do not suffer from this effect, and
appear to produce a smooth redshift histogram: the photometric models
used by LYFS are based entirely on the CWW spectra extended into the
UV without including the effects of evolution, with the effects of the
\Lya\ forest and 912\AA\ break taken into account. However, the
galaxies at $z\approx3$ are predicted to have
$U_{300}-V_{450}\approx5$ and $V_{450}-R{606}\approx1.0$ by LYFS,
which is considerably redder in $U_{300}$ than those detected by
\cite{StHDF96} which have $U_{300}-V_{450}\approx2$ and
$V_{450}-R_{606}\approx0.5$, and therefore casts doubt onto their
photometric models.

To test just how much brighter a distant cluster would have to be to
be clearly evident, we increased the magnitudes of every member of the
simulated cluster at $z=2.0$ until the cluster just became evident in
the images. We emphasize that significantly more than two magnitudes
of brightening would be required to make the cluster stand out from
the field. We stressed that this is far more than the largest expected
error in the brightness of these simulations.

\begin{figure}
\begin{center}
\epsfig{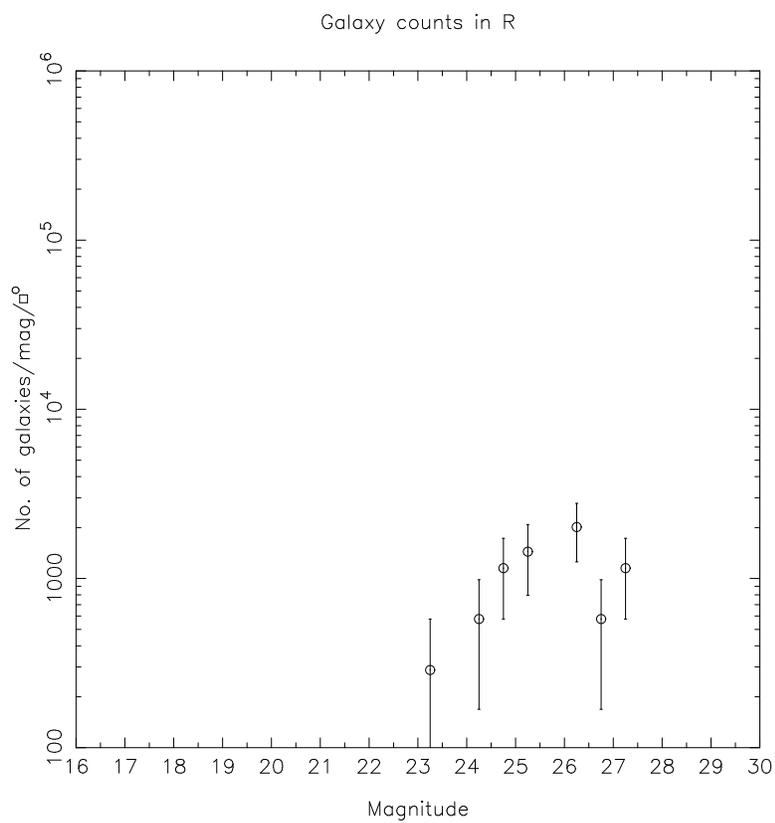}
\caption{$R$ magnitude distribution for the simulated cluster at $z=3.0$
\label{fig:rmag.zeq3.0}
}
\end{center}
\end{figure}

At $z\gtrsim2.3$, the 912\AA\ break starts to extinguish a $U$--band
observation, and by $z\approx3$ provides a clear colour signature,
ie~$U-G>2.0$. This feature has been successfully used to detect
high--redshift galaxies both with custom filters and in the HDF
(eg~\cite{StII}, \cite{StHDF96}). Most of the galaxies previously
detected at $z\approx3$ have $R\gtrsim24.5$ \cite{SGP96}, which is
consistent with the brightest members of the simulated cluster at the
same redshift (Figure~\ref{fig:rmag.zeq3.0}). Intriguingly, while the
photometric redshift estimations do peak at around their expected
value, there is also a significant proportion which are mistaken for
low--redshift galaxies (again, see
Figures~\ref{fig:modalzexp}~\&~\ref{fig:meanzexp}). From the 123
simulated galaxies with colours consistent with $2.5<z<3.0$ and
measured $R<26.0$, 52 are identified as having $z>2.5$ and 70 are
identified as low redshift galaxies with $z<0.8$. If this mimics the
real situation for detecting $z\approx3.0$ galaxies, then it suggests
that a significant number of high--redshift galaxies are missed if
redshifts are estimated using photometric techniques. Note also that
the limiting magnitudes of \cite{StII} etc. are similar to those of
the PC1643 images.

To investigate the efficiency of selecting $z\approx3$ galaxies using
the colour--criteria described in Paper I, we concatenated
the simulated catalogues with $2.5<z<3.5$ and selected objects with
$R<25.5$ and $G>2\sigma$. While the simulations were carried out to
mimic a cluster of galaxies, there is negligible overlap between the
individual simulated galaxies in each catalogue. Therefore, while
there is clustering of the galaxies in the field, it is important to
note that the photometric spread and the results are similar to that
which would be obtained for a random uniform spatial distribution.

\begin{figure}
\begin{center}
\epsfig{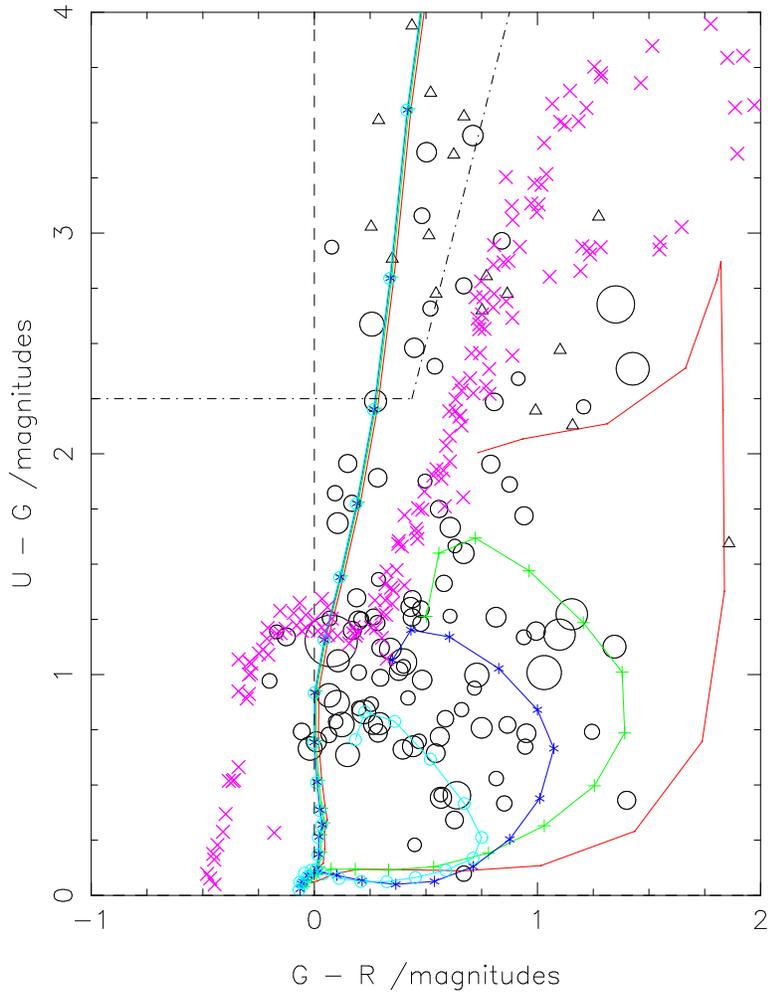}
\caption{$U-G$ vs $G-R$ colour--colour graph for simulated galaxies
with $2.5<z<3.5$. All objects shown have measured $R<25.5$ and
$G>2$--$\sigma$. The triangles denote 1--$\sigma$ lower limits in
$U-G$. The cross are stars taken from the Gunn \& Stryker database,
and the dot--dash line is the bound for selecting high redshift
candidates. The tracks plotted on this graph are the model colours for
galaxies at various redshifts: the point at $U-G=2.2$ and $G-R=0.3$ is
at $z=3.0$, and points at larger $U-G$ are at higher redshift in steps
of $\Delta z=0.1$.
\label{fig:2.5to3.5uggr}
}
\end{center}
\end{figure}

\begin{figure}
\begin{center}
\epsfig{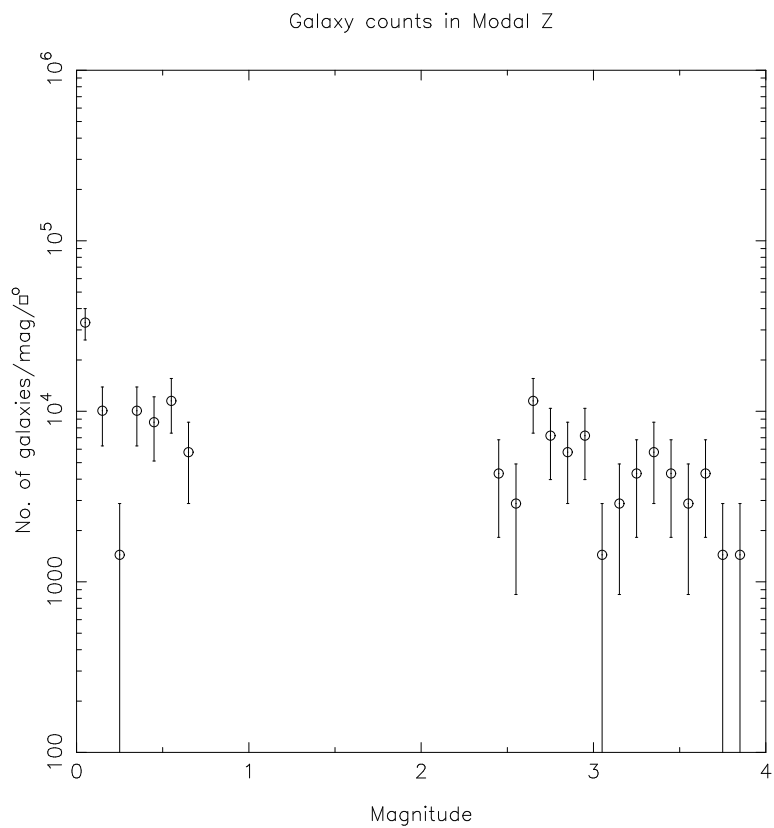}
\caption{Histogram of the photometric redshift distribution measured from a
simulated catalogue containing objects with simulated colours
consistent with galaxies at $2.5<z<3.5$.
\label{fig:2.5to3.5hist}
}
\end{center}
\end{figure}

Figure~\ref{fig:2.5to3.5uggr} shows the measured $U-G$ vs $G-R$
colours of these 104 objects. We note that the simulated objects are
scattered from the simulated colours to such an extent that only 12 of
the objects show measured $U-G$ and $G-R$ colours consistent with
objects at $z\approx3$. Of these same 104 objects, 48 are identified
as being at $z>2$ by the photometric redshift estimator (see
Figure~\ref{fig:2.5to3.5hist}). Even if we take just the catalogue for
the simulated cluster at $z=3.0$, of the 12 objects with $R<25.5$ and
$G>2\sigma$, only two meet the colour--criteria, whereas eight are
identified as having estimated $z>2$. Therefore, five--colour
photometric redshifts appear to provide a more effective method of
detecting high--redshift galaxies than the three--colour $UGR$ method
used by Steidel et al. This is not surprising since the extra
photometric measurements provide extra constraints, which reduce the
effects of individual photometric measurement errors.

\begin{figure}
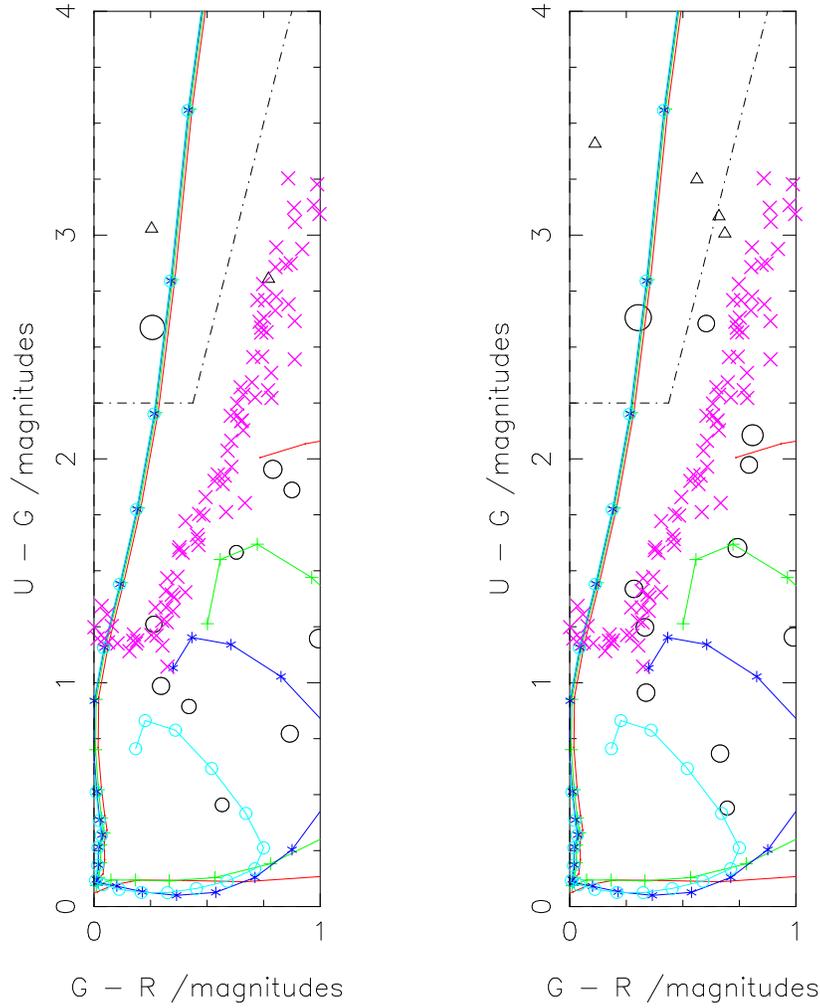

\begin{center}
\begin{minipage}[hp]{0.5\linewidth}
\epsfig{figure=figure59.ps,angle=270,width=0.7\linewidth}
\end{minipage}\hfill
\begin{minipage}[hp]{0.5\linewidth}
\epsfig{figure=figure60.ps,angle=270,width=0.7\linewidth}
\end{minipage}
\caption{Comparison of the $U-G$ vs $G-R$ colours between simulations
with galaxies with angular sizes $\approx10$kpc (left) and galaxies
with angular sizes of $\approx2.5$kpc (right). The dashed line denotes
the limit of the colour--criteria used to select Ly--break
galaxies. All objects plotted have $R<25.5$ and $G>2\sigma$: circles
represent $>1\sigma$ detections in $U$,$G$ and $R$; triangles denote
1--$\sigma$ lower limits for $U-G$.
\label{fig:z=3.0comp}
}
\end{center}
\end{figure}

Comparing the angular sizes of the simulated objects against Ly-break
galaxies observed in the Hubble Deep Field suggests that these
galaxies should be about 2--3kpc in size (\cite{MD97}), which is
considerably smaller than the sizes used in this simulation, which are
$\approx 10$kpc in size. We repeated the $z=3.0$ simulation to assess
whether the angular sizes used had a significant effect, with the
angular sizes of the simulated galaxies reduced by a factor of
four. This small angular size simulation had 15 objects with $R<25.5$
and $G>2\sigma$, of which three fulfilled the colour--criteria,
whereas using photometric redshifts, 8 were identified at
$z>2$. Figure~\ref{fig:z=3.0comp} compares the $U-G$ vs $G-R$ colours
for the original and revised simulation.

The increase in numbers of galaxies detected is due to the increase in
the peak flux of galaxies due to smaller half--light radii -- this
makes it easier for FOCAS to detect these galaxies as their peak
fluxes are more likely to be greater than the $3\sigma$ detection
threshold. It is important to note that decreasing the angular size of
the simulated galaxies has not had a significant effect on the
fraction of galaxies which fulfill the colour criteria for $z\approx3$
galaxies: one might expect reducing the angular size to reduce the
level of confusion observed in these simulations; however, while the
peak flux increases and the area covered by the object is decreased
for all objects, this also promotes simulated objects which were
previously too faint to be detected above the detection threshold and
it is these faint objects which are most likely to be significantly
affected by confusion with the many real faint objects in the
field. The effect of confusion on objects with faint (ie $R>24$)
magnitudes is to ``average'' the simulated object's colours with that
of the confusing object, with the result that a large proportion of
the sample is moved towards the average colours of the real objects in
the field, ie $U-G\approx1.2$ and $G-R\approx1$.

\sloppy{From these results, it appears that the colour--criteria of
\cite{StIII95} is missing a large fraction, 5 out of 6, of the real
population of $z\approx3$ galaxies. This is the result of confusion
due to the $\approx1''$ seeing in the images; one would therefore
expect to find considerably greater numbers of Ly--break galaxies in
the HDF. There are 26 galaxies in the HDF which have spectroscopic
redshifts $z>2$ (\cite{MD97}), of which 12 have $2.8<z<3.5$ and
$V_{606}<25.5$ (this is consistent with the spectroscopic redshifts of
Ly--break candidates; see \cite{SGP96} and the faint magnitude
limit). The area of the Hubble Deep Field is 4.5 arcmin$^2$ and one
would expect, given the published surface--density of ``robust''
Ly--break candidates of $\approx0.7$ arcmin$^{-2}$ (\cite{SAD98}) to find
only 3 Ly--break candidates in the HDF. Because the HDF covers such a
small field of view, there may be significant differences in the
numbers of Ly--break galaxies between adjacent fields, but it is
\emph{striking} that the factor of apparent overdensity of Ly-break
galaxies seen in the HDF is equivalent to the factor of Ly-break
galaxies lost to confusion in the ground--based images.}

In \cite{MD97}, the luminosity function of $z\approx3$ galaxies is
illustrated, but the HDF count of $z\approx3$ galaxies have been
renormalised to the numbers seen in the ground--based images. We
suggest that the surface density seen in the HDF is a better estimate
of the real surface density of $z\approx3$ galaxies, and there are
approximately 6 times more $z\approx3$ galaxies than currently
thought. This has important consequences for the understanding of the
star--formation history of the universe and would suggest that the
peak rate of star--formation occurred at an earlier epoch than
currently thought (see \cite{MPD98}).

To definitively identify the majority of Ly--break candidates and
determine the real surface--density of these galaxies, observations
with much improved seeing will be necessary; either by further HST
observations to increase the area of sky observed and thereby reduce
the effects of cosmic variance, or by deep ground--based observations
with adaptic optics to reduce the point--spread--function in the
images. The effects of confusion can be quantified using similar
simulations to those demonstrated here.

\section{Conclusions}

In using the multicolour photometric redshift technique on the objects found in
the deep imaging of PC1643+4631, we have shown the following.

\begin{enumerate}[(1)]

\item{The technique is efficient at identifying objects at
$z\lesssim1$ where the photometric errors are small, ie at
$R\lesssim24$, because the $\chi^2$ fitting is well constrained because
the colours change strongly with redshift.}

\item{These photometric redshifts are in broad agreement ($\Delta
z\approx0.15$) with the 4 spectroscopic redshifts in the range
$0.62<z<0.81$ in the PC1643+4631 field, whereas HR10 at $z=1.44$ has a
colour--estimated redshift of $z=0.65$, indicating that either
significant reddening can distort the redshift determination or that
at $1<z<2$ the photometric technique fails.}

\item{Ground--based morphological classification is possible on low--redshift
($z\lesssim0.7$) objects with high signal--to--noise photometry (ie
$R<24$ in the deep PC1643 images), although this work does not
investigate its accuracy, due to the difficulties in obtaining secure
morphological classifications for a significant number of galaxies in
this field by manual or automatic means.}

\item{There are degeneracies between the morphological type and
redshift in objects with $1.0\lesssim z \lesssim 1.6$, due to the
models with different morphologies having similar colours at differing
redshifts.}

\item{No constraints can be placed on the morphology of high redshift
($z\gtrsim3$) galaxies using these models since there is no
significant difference between the colours of the model galaxies with
different morphologies at $z\gtrsim2$.}

\item{46 Ly-break candidates have been identified in the PC1643 images
with $2.8<z<3.5$, $R<25.5$ and detected at $>2\sigma$ in $G$; the
distribution of these candidates still shows an underdensity mid--way
between the quasars.}

\item{The histogram of photometric redshift estimates shows two peaks,
at $z\approx0.7$ and $z\approx2.3$, and few objects with $1.0<z<2.0$.
This is probably an artifact of the photometric redshift technique,
arising from the close--spacing of the model colours at these
redshifts, and the histogram is therefore not indicative of the real
distribution of redshifts.}

\item{By calculating both modal and mean photometric redshifts, as
well as a standard deviation for the likelihood distribution, for each
object, ambiguous photometric redshift estimates can easily be flagged
using $|\textrm{modal } z - \textrm{mean } z|>0.2$, giving an
advantage over previous modal--only methods.}

\item{There is evidence of clustering in the field at low
($z\lesssim0.9$) redshift which is consistent with other published
results.}

\newcounter{enumcurr}
\setcounter{enumcurr}{\value{enumi}}

\end{enumerate}

\noindent We have carried out simulations to examine the ability of
the photometric redshift technique to accurately recover simulated
galaxies at various redshifts. These show the following.

\begin{enumerate}[(1)]

\setcounter{enumi}{9}

\item{For simulated galaxies at $0.1<z<0.3$, faint E/S0 galaxies are
mistakenly identified as $z\gtrsim3$ galaxies because of the red $U-G$
colours. An E/S0 with $R\approx24$ will often have no detected flux in
$U$, and be mistaken for a Ly--break candidate.}

\item{For simulated galaxies at $0.1<z<1.0$, the average estimate of
redshift is consistent with the simulated redshift, but there is a
spread of $\Delta z\approx 0.2$ in the distribution. This is partly
due to the photometric errors causing the colours of the object to
move away from the simulated colours, but also due to contamination of
the object colours by other objects in the field.}

\item{For simulated galaxies at $1.0<z<2$, the photometric redshift is
unreliable and misclassifies these galaxies to lower ($z\approx0.6$)
and higher ($z\approx2.2$) redshifts, resulting in peaks in the
photometric redshift distribution which are not representative of the
real redshift distribution. This is a fundamental problem with
photometric redshift estimates based only on optical imaging. A
similar problem is seen in several publications in the literature but
has not been appreciated (see \cite{GH96} and \cite{SLY97}, who both
have a decrease in numbers of galaxies between $1<z<2$ but considered
it a real observation), which has resulted in erroneous conclusions
being drawn.}

\item{For simulated galaxies at $z>2.5$, the photometric technique is
effective in detecting about half the sample as being at $z>2.5$, with
the rest of the sample being mistaken for low redshift galaxies.}

\item{Photometric estimates may prove effective in selecting
high redshift galaxies at $z\approx3$, where the $U-G>2$ colours and
fairly flat spectrum in $G$,$V$,$R$ and $I$ provide additional
constraints to the three colour $U$,$G$,$R$ method currently used for
successfully selecting $z\approx3$ galaxies.}

\item{The scatter of the simulated galaxies away from their simulated
colours is two to three times greater than that expected from the
photometric errors alone and arises as a result of confusion between
the simulated galaxies and the real objects in the field. Since the
amount of confusion is directly related to the resolution of the
images, the WHT images will be far more affected by confusion than
equivalently deep HST images.}

\item{Because of the effects of confusion, the current published
estimates of the surface densities of Ly--break galaxies based on
ground--based imaging are under--estimates by a factor of $\approx6$;
such an increase in the surface density of these galaxies is
consistent with the surface densities seen in the HDF.}

\item{Any third image of a single quasar causing the PC1643 pair is
extremely likely to be confused and therefore unrecognisable.}

\item{Deep ground--based CCD imaging without adaptive optics is
\emph{inevitably confusion--limited} at $R>26.0$.  Imaging with
smaller point--spread--functions is the \emph{only} way to alleviate
this.}

\end{enumerate}

Simulations have been carried out to investigate the difficulties in
identifying a cluster at various redshifts in the PC1643 field.
These show the following.

\begin{enumerate}[(1)]

\setcounter{enumi}{18}

\item{Even with our many optical colours, it is difficult to visually
identify a cluster of galaxies at $z\approx0.4$ against the other
objects in any deep field. A cluster at $z=1$ with galaxy luminosities
similar to those seen locally would be lost against the other objects
in the field.}

\item{Even a very rich cluster at $z\approx1.0$ would be impossible to
distinguish from the field on the basis of this $U$, $G$, $V$, $R$ and
$I$ imaging and hence the evident absence of a cluster in these
$1.1''$ resolution images is entirely consistent with a distant
cluster producing the CMB decrement seen.}

\item{A cluster at $z=1.0$ in which each member was 1 magnitude
brighter than expected would still be difficult to detect; however, if
the members were more than 2 magnitudes brighter, the cluster would be
evident in the images.}

\end{enumerate}

\section{Acknowledgements}
GC acknowledges a PPARC Postdoctoral Research
Fellowship. TH acknowledges a PPARC
studentship. We thank Steve Rawlings for obtaining the redshift of
galaxy 4 in Table 1.

\bibliographystyle{apalike}
\bibliography{general}

\clearpage

\begin{table}
\begin{center}
\begin{tabular}{ccc}\hline
 		& Spectroscopic 	& Photometric \\
Object  	& redshift		& redshift \\ \hline \hline

1		& $0.659\pm0.002$	& $0.55\pm0.03$\\
2		& $0.670\pm0.002$	& $0.57\pm0.04$\\
3		& $0.670\pm0.002$	& $0.52\pm0.05$\\
4		& $0.810\pm0.003$	& $0.62\pm0.08$\\
HR10		& $1.44\pm0.01$		& $0.65\pm0.08$\\
\hline
\end{tabular}
\caption{Comparison of spectroscopic and photometric redshifts in
PC1643\label{tab:compsp}}
\end{center}
\end{table}

\clearpage

\clearpage
\clearpage

\clearpage

\clearpage

\clearpage

\clearpage

\clearpage

\clearpage

\clearpage

\clearpage

\clearpage

\clearpage

\clearpage

\clearpage

\clearpage
\begin{figure}
\begin{center}
\begin{minipage}[hp]{0.5\linewidth}
\epsfig{figure=figure27.ps,angle=270,width=0.85\linewidth}
\end{minipage}\hfill
\begin{minipage}[hp]{0.5\linewidth}
\epsfig{figure=figure28.ps,angle=270,width=0.85\linewidth}
\end{minipage}
\caption{Two low--redshift Sbc--type galaxies
\label{fig:lowzSbc}
}
\end{center}
\end{figure}

\clearpage
\begin{figure}
\begin{center}
\begin{minipage}[hp]{0.5\linewidth}
\epsfig{figure=figure29.ps,angle=270,width=0.85\linewidth}
\end{minipage}\hfill
\begin{minipage}[hp]{0.5\linewidth}
\epsfig{figure=figure30.ps,angle=270,width=0.85\linewidth}
\end{minipage}
\caption{Two low--redshift Scd--type galaxies
\label{fig:lowzScd}
}
\end{center}
\end{figure}

\clearpage

\clearpage

\clearpage

\clearpage

\clearpage

\clearpage

\clearpage

\clearpage

\clearpage
\begin{figure}
\begin{center}

This figure is avaliable at ftp.mrao.cam.ac.uk:/pub/PC1643/paper2.figure50.eps

\caption{Colour images of the simulated cluster at $z=0.4$, $z=0.6$
and $z=0.8$. Images on the left are on a blank background; those on
the right are superimposed on the PC1643 field. All these images use
the same colouring scheme used in Cotter et al., 1998.
\label{fig:0.2-0.8.images}
}
\end{center}
\end{figure}

\clearpage
\begin{figure}
\begin{center}

This figure is avaliable at ftp.mrao.cam.ac.uk:/pub/PC1643/paper2.figure51.eps

\caption{Colour images of the simulated cluster at $z=1.0$, $z=2.0$
and $z=4.0$. Images on the left are on a blank background;
those on the right are superimposed on the PC1643 field. All these
images use the same colouring scheme used in
Cotter et al., 1998.
\label{fig:1.0-4.0.images}
}
\end{center}
\end{figure}

\clearpage
\begin{figure}
\begin{center}
\epsfig{figure=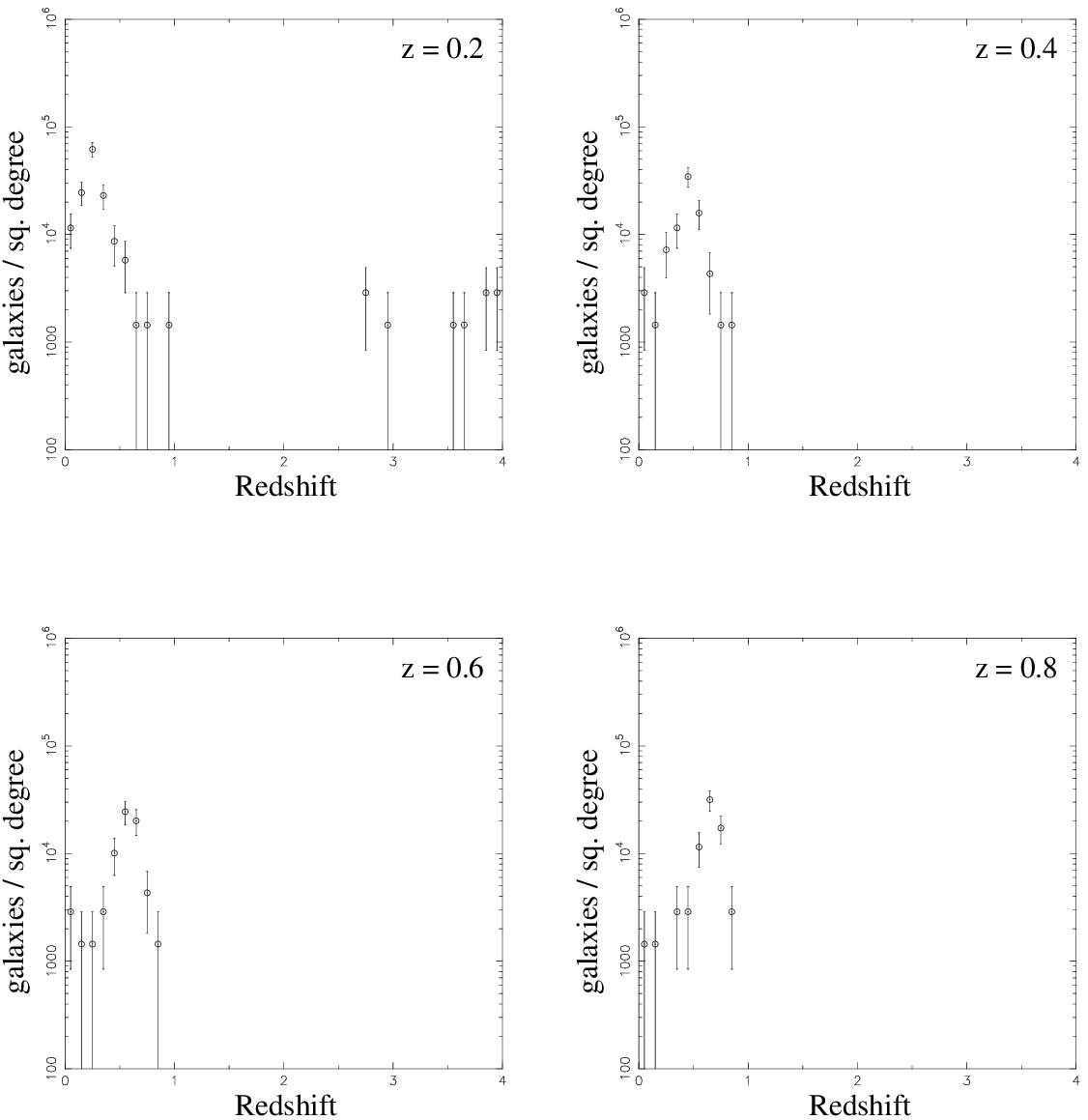,angle=0,width=0.95\linewidth}
\caption{The histogram of modal redshifts as estimated from
photometric measurements of the simulated galaxies. These graphs
illustrate the effect of attempting to recover the simulated galaxies
from the PC1643 field using FOCAS at $z=0.2$, $z=0.4$, $z=0.6$ and
$z=0.8$. Only simulated galaxies with measured $R<26.0$ are shown.
\label{fig::0.2-0.8.graphs}
}
\end{center}
\end{figure}

\clearpage
\begin{figure}
\begin{center}
\epsfig{figure=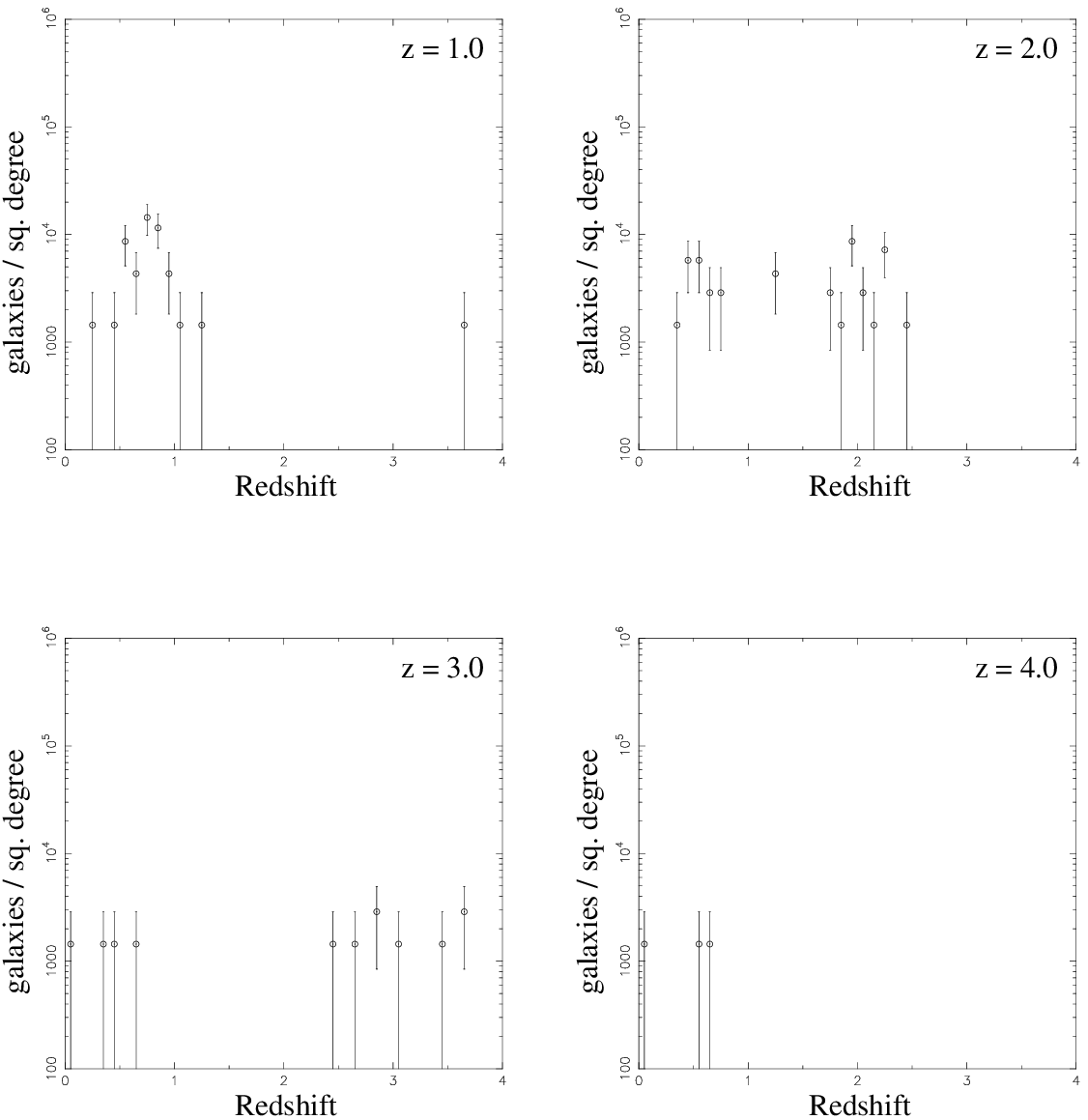,angle=0,width=0.95\linewidth}
\caption{The histogram of modal redshifts as estimated from
photometric measurements of the simulated galaxies. These graphs
illustrate the effect of attempting to recover the simulated galaxies
from the PC1643 field using FOCAS at $z=1.0$, $z=2.0$, $z=3.0$ and
$z=4.0$. Only simulated galaxies with measured $R<26.0$ are shown.
\label{fig:1.0-4.0.graphs}
}
\end{center}
\end{figure}

\clearpage

\clearpage

\clearpage

\clearpage

\clearpage

\clearpage

\end{document}

%% file: journals.tex
%
%
%


\let\jnlstyle=\rm
\def\refjnl#1{{\jnlstyle#1}}

\def\aj{\refjnl{AJ}}                   
\def\araa{\refjnl{ARA\&A}}             
\def\apj{\refjnl{ApJ}}                 
\def\apjl{\refjnl{ApJ}}                
\def\apjs{\refjnl{ApJS}}               
\def\ao{\refjnl{Appl.~Opt.}}           
\def\apss{\refjnl{Ap\&SS}}             
\def\aap{\refjnl{A\&A}}                
\def\aapr{\refjnl{A\&A~Rev.}}          
\def\aaps{\refjnl{A\&AS}}              
\def\azh{\refjnl{AZh}}                 
\def\baas{\refjnl{BAAS}}               
\def\jrasc{\refjnl{JRASC}}             
\def\memras{\refjnl{MmRAS}}            
\def\mnras{\refjnl{MNRAS}}             
\def\pra{\refjnl{Phys.~Rev.~A}}        
\def\prb{\refjnl{Phys.~Rev.~B}}        
\def\prc{\refjnl{Phys.~Rev.~C}}        
\def\prd{\refjnl{Phys.~Rev.~D}}        
\def\pre{\refjnl{Phys.~Rev.~E}}        
\def\prl{\refjnl{Phys.~Rev.~Lett.}}    
\def\pasp{\refjnl{PASP}}               
\def\pasj{\refjnl{PASJ}}               
\def\qjras{\refjnl{QJRAS}}             
\def\skytel{\refjnl{S\&T}}             
\def\solphys{\refjnl{Sol.~Phys.}}      
\def\sovast{\refjnl{Soviet~Ast.}}      
\def\ssr{\refjnl{Space~Sci.~Rev.}}     
\def\zap{\refjnl{ZAp}}                 
\def\nat{\refjnl{Nature}}              
\def\iaucirc{\refjnl{IAU~Circ.}}       
\def\aplett{\refjnl{Astrophys.~Lett.}} 
\def\apspr{\refjnl{Astrophys.~Space~Phys.~Res.}}
\def\bain{\refjnl{Bull.~Astron.~Inst.~Netherlands}} 
\def\fcp{\refjnl{Fund.~Cosmic~Phys.}}  
\def\gca{\refjnl{Geochim.~Cosmochim.~Acta}}   
\def\grl{\refjnl{Geophys.~Res.~Lett.}} 
\def\jcp{\refjnl{J.~Chem.~Phys.}}      
\def\jgr{\refjnl{J.~Geophys.~Res.}}    
\def\jqsrt{\refjnl{J.~Quant.~Spec.~Radiat.~Transf.}}
\def\memsai{\refjnl{Mem.~Soc.~Astron.~Italiana}}
\def\nphysa{\refjnl{Nucl.~Phys.~A}}   
\def\physrep{\refjnl{Phys.~Rep.}}   
\def\physscr{\refjnl{Phys.~Scr}}   
\def\planss{\refjnl{Planet.~Space~Sci.}}   
\def\procspie{\refjnl{Proc.~SPIE}}   

\let\astap=\aap
\let\apjlett=\apjl
\let\apjsupp=\apjs
\let\applopt=\ao

%% file: paper2.bbl
\begin{thebibliography}{}

\bibitem[{Adami} et~al., 1998]{ABM98}
{Adami}, C., {Biviano}, A., and {Mazure}, A. (1998).
\newblock Segregations in clusters of galaxies.
\newblock {\em \aap}, 331:439--450.

\bibitem[{Bruzual} and {Charlot}, 1993]{BC93}
{Bruzual}, G.~A. and {Charlot}, S. (1993).
\newblock Spectral evolution of stellar populations using isochrone synthesis.
\newblock {\em \apj}, 405:538--553.

\bibitem[{Coleman} et~al., 1980]{CWW}
{Coleman}, G.~D., {Wu}, C.~C., and {Weedman}, D.~W. (1980).
\newblock Colors and magnitudes predicted for high redshift galaxies.
\newblock {\em \apjs}, 43:393--416.

\bibitem[Dickinson, 1997]{MD97}
Dickinson, M. (1997).
\newblock {C}olor-{S}elected {H}igh {R}edshift {G}alaxies and the {HDF}.
\newblock In Livio, M.~Fall, S. and Madau, P., editors, {\em {T}he {H}ubble
  {D}eep {F}ield}, {STScI} {S}ymposium.

\bibitem[{Dressler} et~al., 1997]{DOC97}
{Dressler}, A., {Oemler}, Augustus, J., {Couch}, W.~J., {Smail}, I., {Ellis},
  R.~S., {Barger}, A., {Butcher}, H., {Poggianti}, B.~M., and {Sharples}, R.~M.
  (1997).
\newblock Evolution since z = 0.5 of the {M}orphology-{D}ensity {R}elation for
  {C}lusters of {G}alaxies.
\newblock {\em \apj}, 490:577+.

\bibitem[{Graham} and {Dey}, 1996]{GD96}
{Graham}, J.~R. and {Dey}, A. (1996).
\newblock The redshift of an {E}xtremely {R}ed {O}bject and the nature of the
  very red galaxy population.
\newblock {\em \apj}, 471:720+.

\bibitem[{Gwyn}, 1998]{SG98}
{Gwyn}, S. D.~J. (1998).
\newblock Photometric redshifts.
\newblock {W}orld {W}ide {W}eb page, University of {V}ictoria, {C}anada.
\newblock http://astrowww.phys.uvic.ca/grads/gwyn/pz/index.html.

\bibitem[{Gwyn} and {Hartwick}, 1996]{GH96}
{Gwyn}, S. D.~J. and {Hartwick}, F. D.~A. (1996).
\newblock The redshift distribution and luminosity functions of galaxies in the
  {H}ubble {D}eep {F}ield.
\newblock {\em \apjl}, 468:L77--+.

\bibitem[{Hogg} et~al., 1998]{HCB98}
{Hogg}, D.~W., {Cohen}, J.~G., {Blandford}, R., {Gwyn}, S. D.~J., {Hartwick},
  F. D.~A., {Mobasher}, B., {Mazzei}, P., {Sawicki}, M., {Lin}, H., {Yee}, H.
  K.~C., {Connolly}, A.~J., {Brunner}, R.~J., {Csabai}, I., {Dickinson}, M.,
  {Subbarao}, M.~U., {Szalay}, A.~S., {Fern\'andez-Soto}, A., {Lanzetta},
  K.~M., and {Yahil}, A. (1998).
\newblock A blind test of photometric redshift prediction.
\newblock {\em \aj}, 115:1418--1422.

\bibitem[{Hu} and {Ridgway}, 1994]{HuR94}
{Hu}, E.~M. and {Ridgway}, S.~E. (1994).
\newblock Two extremely red galaxies.
\newblock {\em \aj}, 107:1303--1306.

\bibitem[{Kinney} et~al., 1993]{KBC93}
{Kinney}, A.~L., {Bohlin}, R.~C., {Calzetti}, D., {Panagia}, N., and {Wyse}, R.
  F.~G. (1993).
\newblock An {A}tlas of {U}ltraviolet {S}pectra of {S}tar-{F}orming {G}alaxies.
\newblock {\em \apjs}, 86:5--93.

\bibitem[{Lanzetta} et~al., 1996]{LYF96}
{Lanzetta}, K.~M., {Yahil}, A., and {Fernandez-Soto}, A. (1996).
\newblock Star-forming galaxies at very high redshifts.
\newblock {\em \nat}, 381:759--763.

\bibitem[Lilly et~al., 1995]{CFRS1}
Lilly, S., Le~F\`evre, O., Crampton, D., Hammer, F., and Tresse, L. (1995).
\newblock {T}he {C}anada--{F}rance {R}edshift {S}urvey. {I}. {I}ntroduction to
  the survey, photometric catalogues and surface brightness selection effects.
\newblock {\em \apj}, 455:50--59.

\bibitem[{Lilly} et~al., 1995]{LTH95}
{Lilly}, S.~J., {Tresse}, L., {Hammer}, F., {Crampton}, D., and {Le Fevre}, O.
  (1995).
\newblock The {C}anada-{F}rance {R}edshift {S}urvey. {VI}. {E}volution of the
  {G}alaxy {L}uminosity {F}unction to z approximately 1.
\newblock {\em \apj}, 455:108+.

\bibitem[{Madau} et~al., 1998]{MPD98}
{Madau}, P., {Pozzetti}, L., and {Dickinson}, M. (1998).
\newblock The star formation history of field galaxies.
\newblock {\em \apj}, 498:106+.

\bibitem[{Sawicki} et~al., 1997]{SLY97}
{Sawicki}, M.~J., {Lin}, H., and {Yee}, H. K.~C. (1997).
\newblock Evolution of the galaxy population based on photometric redshifts in
  the hubble deep field.
\newblock {\em \aj}, 113:1--12.

\bibitem[{Schechter}, 1976]{Schechter}
{Schechter}, P. (1976).
\newblock An analytic expression for the luminosity function for galaxies.
\newblock {\em \apj}, 203:297--306.

\bibitem[{Steidel} et~al., 1998]{SAD98}
{Steidel}, C.~C., {Adelberger}, K.~L., {Dickinson}, M., {Giavalisco}, M.,
  {Pettini}, M., and {Kellogg}, M. (1998).
\newblock A large structure of galaxies at redshift z approximately 3 and its
  cosmological implications.
\newblock {\em \apj}, 492:428+.

\bibitem[{Steidel} et~al., 1996a]{StHDF96}
{Steidel}, C.~C., {Giavalisco}, M., {Dickinson}, M., and {Adelberger}, K.~L.
  (1996a).
\newblock Spectroscopy of {L}yman break galaxies in the {H}ubble {D}eep
  {F}ield.
\newblock {\em \aj}, 112:352+.

\bibitem[{Steidel} et~al., 1996b]{SGP96}
{Steidel}, C.~C., {Giavalisco}, M., {Pettini}, M., {Dickinson}, M., and
  {Adelberger}, K.~L. (1996b).
\newblock Spectroscopic confirmation of a population of normal star-forming
  galaxies at redshifts z $>$ 3.
\newblock {\em \apjl}, 462:L17--+.

\bibitem[{Steidel} and {Hamilton}, 1993]{StII}
{Steidel}, C.~C. and {Hamilton}, D. (1993).
\newblock Deep imaging of high redshift {QSO} fields below the {L}yman limit.
  {II} - {N}umber counts and colors of field galaxies.
\newblock {\em \aj}, 105:2017--2030.

\bibitem[{Steidel} et~al., 1995]{StIII95}
{Steidel}, C.~C., {Pettini}, M., and {Hamilton}, D. (1995).
\newblock {L}yman imaging of high-redshift galaxies.{III}.{N}ew observations of
  four {QSO} fields.
\newblock {\em \aj}, 110:2519+.

\bibitem[{Wu} and {Hammer}, 1993]{WH93}
{Wu}, X.-P. and {Hammer}, F. (1993).
\newblock Statistics of lensing by clusters of galaxies. {I} - {G}iant arcs.
\newblock {\em \mnras}, 262:187--203.

\end{thebibliography}
